\documentclass[a4paper,fleqn,usenatbib,useAMS]{mnras}

\usepackage{array, multirow, graphicx}	% Including figure files
\usepackage{amsmath}	% Advanced maths commands
\usepackage{amssymb}	% Extra maths symbols
\usepackage{multicol}        % Multi-column entries in tables
\usepackage{pdflscape}	% Landscape pages
\usepackage{subfigure}
\usepackage{wasysym}

\usepackage[T1]{fontenc}
\usepackage{ae,aecompl}

\usepackage{newtxtext,newtxmath}
\usepackage{tabularx}

\title[Polarization measurements of G29-38]{Polarization measurements of the polluted white dwarf G29-38}

\author[D. V. Cotton \textit{et al.}]{Daniel V. Cotton$^{1,2,3}$\thanks{Contact e-mail: \href{mailto:Daniel.Cotton@anu.edu.au}{Daniel.Cotton@anu.edu.au}}, Jeremy Bailey$^{4}$, J. E. Pringle$^{5}$,  William B. Sparks$^{6,7}$,
\newauthor Ted von Hippel$^{8}$, Jonathan P. Marshall$^{9}$\\
\\
$^1$Anglo Australian Telescope, Australian National University, 418 Observatory Road, Coonabarabran, NSW 2357, Australia.\\
$^2$Western Sydney University, Locked Bag 1797, Penrith-South DC, NSW 1797, Australia.\\
$^3$Centre for Astrophysics, University of Southern Queensland, Toowoomba, Queensland. 4350. Australia.\\
$^4$School of Physics, UNSW Sydney, New South Wales, 2052, Australia.\\
$^5$Institute of Astronomy, Madingley Road, Cambridge, CB3 0HA, UK.\\
$^6$SETI Institute, 189 Bernardo Avenue, Suite 200, Mountain View, CA 94043, USA.\\
$^7$Space Telescope Science Institute, 3700 San Martin Drive, Baltimore, MD 21218, USA.\\
$^8$Department of Physical Sciences, Embry-Riddle Aeronautical University, 1 Aerospace Blvd, Daytona Beach, FL 32114, USA.\\
$^{9}$Academia Sinica, Institute of Astronomy and Astrophysics, 11F Astronomy-Mathematics Building, NTU/AS campus, No. 1, Section 4, \\
Roosevelt Rd., Taipei 10617, Taiwan.\\
}

\date{Last updated \today; in original form \today}

\pubyear{2020}
\begin{document}
\label{firstpage}
\pagerange{\pageref{firstpage}--\pageref{lastpage}}
\maketitle

\begin{abstract}
    We have made high precision polarimetric observations of the polluted white dwarf \mbox{G29-38} with the HIgh Precision Polarimetric Instrument 2. The observations were made at two different observatories -- using the \mbox{8.1-m} Gemini North Telescope and the \mbox{3.9-m} Anglo Australian Telescope -- and are consistent with each other. After allowing for a small amount of interstellar polarization, the intrinsic linear polarization of the system is found to be 275.3~$\pm$~31.9~parts-per-million at a position angle of 90.8~$\pm$~3.8$^\circ$ in the SDSS $g^{\prime}$ band. We compare the observed polarization with the predictions of circumstellar disc models. The measured polarization is small in the context of the models we develop which only allows us to place limits on disc inclination and Bond albedo for optically thin disc geometries. In this case either the inclination is near face-on or the albedo is small -- likely in the range 0.05 to 0.15 -- which is in line with other debris disc measurements. A preliminary search for the effects of G29-38's pulsations in the polarization signal produced inconsistent results. This may be caused by beating effects, indicate a clumpy dust distribution, or be a consequence of measurement systematics.
\end{abstract}

% Select between one and six entries from the list of approved keywords.
% Don't make up new ones.

\section{Introduction}

White dwarfs, the end state of stellar evolution for stars of $\sim$0.5 -- 8~M$_{\astrosun}$, provide strong evidence for planetary systems orbiting their precursor main sequence stars, as well as their subsequent survival through post-main sequence stellar evolution \citep{Farihi16}. Specifically, 25~per~cent of white dwarfs show metal lines in their atmospheres \citep{Zuckerman03}. For white dwarfs with hydrogen atmospheres, the sinking timescale for metals to leave the atmosphere due to the extreme gravity is only weeks to centuries \citep{Koester06}. These timescales imply that the metals must have arrived recently and may still be arriving. Remnants of the precursor planetary system are the most likely source of these metals. Additionally, around 4~per~cent of these stars exhibit infrared excesses \citep{Rocchetto15, Wilson19}, typically re-radiating approximately 1~per~cent of the stellar flux. This fact, along with infrared spectroscopy showing silicates in emission in these systems \citep{Reach05}, is broadly interpreted to indicate that these stars are actively accreting dust from asteroids and/or comets. Many white dwarfs with excesses have discs with substantial gas components (e.g. \citealp{Farihi12}) which may be associated with dissociated planetesimals within the tidal disruption radius \citep{Wilson14}. With the discovery of transits in the time-series photometry of WD~1145+017 \citep{Vanderburg15, Zhou16}, and optical spectroscopy of WD~J0914+1914 indicating an evaporating giant planet \citep{Gansicke19}, the evidence appears overwhelming that white dwarfs are providing insight into the ubiquity and composition of extrasolar minor bodies. Indeed, observations of these systems have already motivated dynamical studies of the scenarios of their creation, assuming they are the result of inward scattering of planetesimals \citep{Bonsor11, Bonsor15}.

Yet, from an empirical standpoint, the fundamental properties of these systems remain poorly constrained. Of particular concern is whether the circumstellar dust is optically thin or thick and the geometry of its distribution. Without this knowledge it is impossible to accurately calculate the dust masses or circumstellar debris lifetimes in these systems \citep{Jura03, Farihi14, Bonsor17}, which in turn are necessary to constrain their nature and the dynamical mechanism(s) that brought the minor bodies close enough to the host star for what is thought to be tidal disruption \citep{Jura03}, followed by eventual accretion.

G29-38 (ZZ~Psc, WD~2326+049) became the prototype white dwarf with circumstellar dust when its infrared excess was discovered at the NASA InfraRed Telescope Facility some thirty years ago \citep{Zuckerman87}. It is a DA white dwarf with $T_{\rm eff} = 11\,240$ K, $\log g = 8.00$ and is at a distance of $d = 17.5$~pc \citep{Xu18}. This star has a rare yet auspicious combination of properties that should allow us eventually to place some constraints on the geometry and nature of its circumstellar dust. Based on its strong infrared excess ($L_{\rm IR}/L_\star$ = 3.9~per~cent, \citealp{Farihi14}), it has one of the most substantial circumstellar dust reservoirs. It is also a well-studied luminosity variable pulsating in a range of non-radial gravity-modes with periods of 1 to 10 minutes \citep{Kleinman98}. Pioneering observations \citep{Graham90, Patterson91} identified corresponding oscillations in the K-band, which can provide a measurement of the modes as seen from the dust rather than directly by the observer.

Magnetic white dwarfs, those with surface magnetic fields stronger than 10~kG, can have significant linear polarization -- up to several per~cent (e.g. \citealp{West89}). When linear polarization is seen in blind surveys of white dwarfs, it is attributed to magnetism \citep{Zejmo16, Slowikowska18}. However, G29-38 has a magnetic field less than $\sim$100~G \citep{Liebert89}, so it is assumed that a measured polarization will be due to its dust.

Understanding the nature of the dust, and the geometry of the dust distribution, is critical for understanding the nature of the planetary debris progenitor material and the mechanism(s) by which it is initially disrupted, and eventually finds its way to the stellar surface. Such dust is usually assumed to be in the form of a thin disc close to the star, arising from the disruption of a single asteroid. However, if the dust around the star is due to multiple asteroids and their resulting collisions, the geometry of the dust might be more widespread. \citet{Reach09} consider the infrared spectra that would result from a variety of dust distributions.

In this paper we report a measurement of the polarization of optical light from G29-38, which we take to be the result of scattering from the dust that surrounds the object and gives rise to the infrared excess. We also report an attempt to look for variability of that polarization, which could be interpreted as resulting from the known variability of the star.

In Section~\ref{obsres} we report the details of the observations and the results, including making a determination of, and correction for, interstellar polarization in Section~\ref{interstellar}. In Section~\ref{analysis} we analyse the data by first giving a brief overview of how these results might be analysed in Section \ref{basics}; then in Section~\ref{dust} we summarise the assumptions we shall make about the general properties of the dust that we require for modelling the data; and in Section~\ref{models} we present models for the dust distribution and show the implications for the dust properties, given the measured polarization; before looking at a potential complications (Section~\ref{clumpy}). We present a brief discussion in section~\ref{discussion} before presenting our conclusions in Section~\ref{conclusions}.

\section{Observations and results}
\label{obsres}

\subsection{Standard G29-38 observations}
\label{Obs}

We have observed G29-38 (ZZ~Psc, WD~2326+049) with (two copies of) the HIPPI-2 polarimeter \citep{Bailey19b} during two observing runs. The first run was made at Mauna Kea in Hawaii with the Alt-Az 8.1-m Gemini North telescope during early 2018 July (denoted GN). The second run was made at Siding Spring Observatory in Australia with the equatorial 3.9-m Anglo-Australian Telescope (AAT) during 2018 August. HIPPI-2 is designed for an f/16 focus appropriate for the Gemini North observations. For the AAT run, the f/8 Cassegrain focus was used with the addition of a $\times2$ negative achromatic (Barlow) lens. In combination with the selected instrument apertures (4~mm GN, 3.6~mm AAT) this gives on-sky aperture sizes of 6.4 arcsec (GN) and 11.9 arcsec (AAT).

HIPPI-2 is a high-precision optical linear aperture polarimeter based on the proven HIPPI \citep{Bailey15} and Mini-HIPPI \citep{Bailey17} instruments, and is described in full in \citet{Bailey19b}. It has recently been used to study polarization in the binary star system Spica \citep{Bailey19a}, the rapidly rotating star $\alpha$~Oph \citep{Bailey20b}, and the red supergiant Betelgeuse \citep{Cotton20a}. To summarise its performance, in the SDSS $g^{\prime}$ band, in which our observations were made, the instrument has demonstrated precision of better than 4~parts-per-million (ppm). It achieves this through two stages of modulation. The first stage is rapid (500~Hz) and electrically driven in order to beat seeing noise; it is accomplished by a commercial Ferro-electric Liquid Crystal (FLC) modulator. The second stage of modulation is slower, and involves utilising an instrument rotator to achieve four instrument position angles (PA) of 0$^\circ$, 45$^\circ$, 90$^\circ$ and 135$^\circ$ per observation. The redundant angles (90$^\circ$ and 135$^\circ$) are used to cancel out instrumental effects. For the AAT run the FLC used was the MS series polarization rotator from Boulder Non-Linear Systems (BNS). At Gemini North we used an FLC from Meadowlark Optics (ML) with a design wavelength of 500~nm. The performance of each unit has been characterised by both laboratory and on-sky measurements, and is described in detail in \citet{Bailey19b}. 

HIPPI-2 uses a Wollaston prism as an analyser, whence the light is fed into twin compact photomultiplier tubes (PMTs) used as detectors. For these observations we utilised two different pairs of Hamamatsu H10720-210 modules which have ultrabialkali photocathodes \citep{Nakamura10} providing a quantum efficiency of 43~per~cent at 400~nm. In combination with the atmosphere, telescope and instrument optics this results in a mean effective wavelength for G29-38 observations of 451.7~nm (AAT) and 463.6~nm (Gemini North), as calculated by a bandpass model -- the difference is mainly down to the reflectance profile of Gemini North's silver coated mirror compared to the AAT's aluminium mirror. The bandpass model also takes account of modulation efficiency of the instrument, the inverse of which must be multiplied by the raw Stokes parameter measurements to determine the true polarization. The mean modulation efficiency was significantly impacted by the performance characteristics of the different modulators; for AAT observations it was 63.5~per~cent, on Gemini the figure was 93.4~per~cent. The AAT run was the final one for the BNS modulator, the performance of which drifted over time, accounting for the comparatively low modulation efficiency. The bandpass model requires a source spectrum. For this purpose we chose to use the synthetic white dwarf spectra of \citet{Koester10}. \citet{Koester09} gives $T_{\rm eff}=$11\,485~K and $\log(g)=$8.071 for G29-38; the two nearest models have $T_{\rm eff}=$11\,000~K and 12\,000~K. We therefore ran the bandpass calculations with both models separately and took the mean.

The telescope polarization (TP) and position angle (PA) calibrations rely respectively on the observation of low and high polarization standard stars. These observations and the associated calibrations for the two runs have both been reported previously \citep{Bailey19b}. On the AAT the TP was 13.6~$\pm$~1.1~ppm at a position angle of 80.9~$\pm$~2.2$^\circ$. On an Alt-Az telescope the TP varies with parallactic angle for any given Stokes parameter. The TP at Gemini North is very large (around 2,050~ppm in the SDSS $g^{\prime}$ band), and highly wavelength dependant, meaning that it also varies significantly with airmass. A model describing the TP at Gemini North was developed, and is presented elsewhere \citep{Bailey19b}. The uncertainties surrounding the TP limit the achievable precision (in $g^{\prime}$) at Gemini North to 24.9~ppm.

It is normal procedure with HIPPI-2 to make a measurement of the sky at each PA immediately before or after each measurement of the object. At Gemini North this was done in the normal way with alternating trailing and leading skies. On the AAT for observations of G29-38 we took skies that bracketed every measurement and averaged them before making the subtraction. This precaution against changing sky conditions was taken because, at $m_B=13.17$~mag, G29-38 is one of the faintest objects so far observed with a HIPPI-class polarimeter.

\begin{figure*}
\centering
\includegraphics[clip, trim={1.8cm 1.5cm 2cm, 1cm}, width=17.5cm]{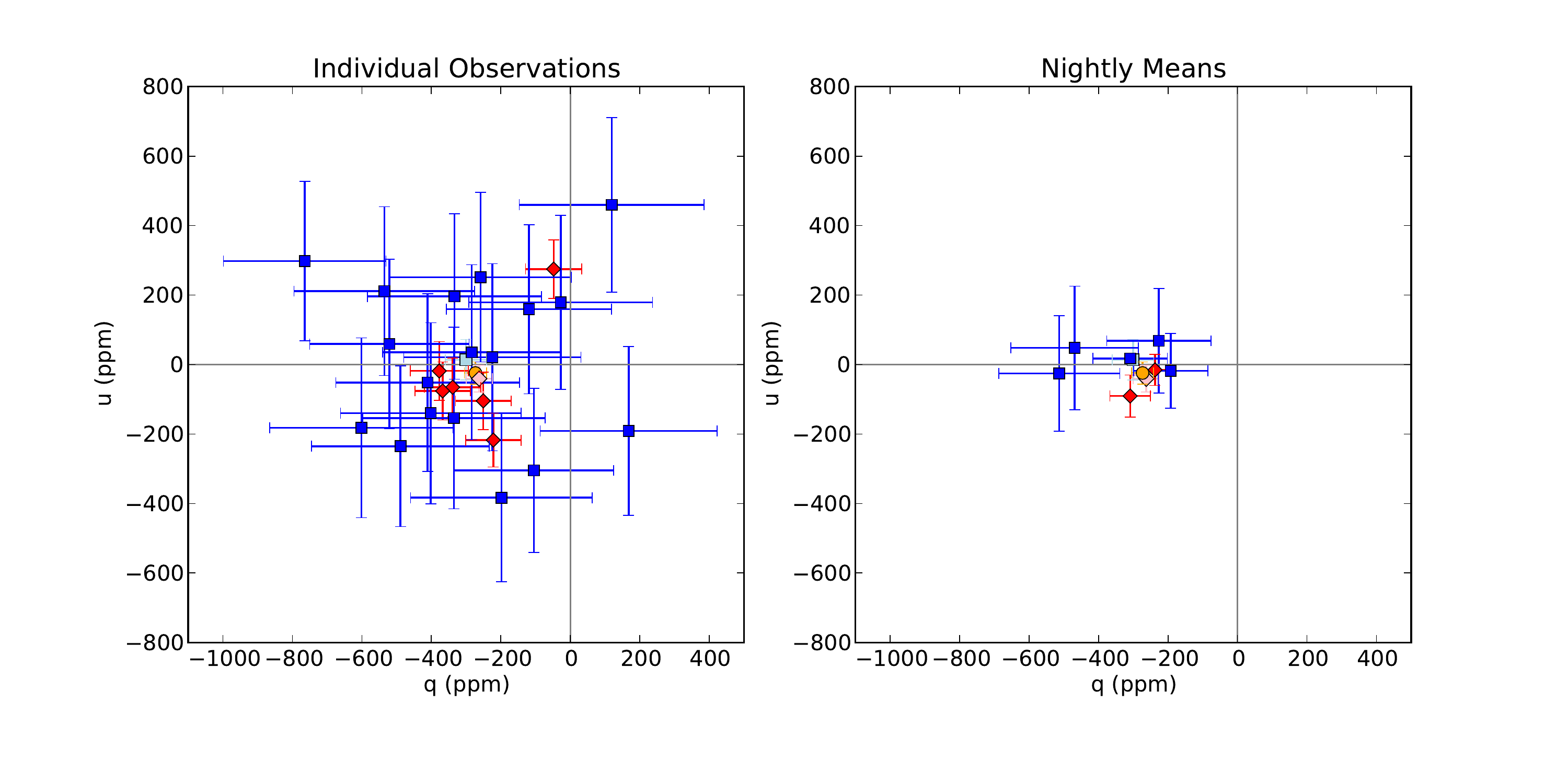}
\caption{Q-U diagrams of HIPPI-2 observations of G29-38. Gemini North observations are shown as red diamonds, AAT observations as blue squares. The left hand panel shows the individual observations, the right hand panel the nightly means. In both panels the mean of all observations is shown as an orange circle, and the means of the GN and AAT runs shown as a pink diamond and light blue square (mostly obscured in the right hand panel) respectively.}
\label{fig:pol_obs}
\end{figure*}

The individual HIPPI-2 observations are given in Table \ref{tab:pol_obs} and presented in the form of a Q-U diagram in Figure \ref{fig:pol_obs}, along with nightly means, run means, and the mean of both runs\footnote{We use $Q$ and $U$ to denote the linear polarization Stokes parameters, and $q$ and $u$ the normalised forms, i.e. $q=Q/I$, $u=U/I$.}. In all cases the means are error weighted. The errors on the individual observations are fairly large, but these appear fairly consistent. This can be seen most clearly in the left-hand panel of Figure \ref{fig:pol_obs}, where only one of the individual Gemini North observations deviates by more than 2$\sigma$ from the mean. More meaningfully, the nightly means are in good agreement within the formal error, as are the run means. The latter, in particular, gives confidence in the result, and implies stability in the system polarization on the scale of days to weeks. Combining all the data gives $p=$274.3$\pm$31.8~ppm at a PA of 92.5~$\pm$~3.3$^\circ$, a clear detection at a significance at 8.6$\sigma$.

\begin{table*}
\centering
\caption{HIPPI-2 Observations of G29-38}
\tabcolsep 6.2 pt
\begin{tabular}{cllllrrrrrrr}
\hline
\hline
&Run & \multicolumn{3}{c}{Observation Time}  & \multicolumn{1}{c}{Exp.}  & \multicolumn{1}{c}{$\lambda_{\rm eff}$} & \multicolumn{1}{c}{Eff.}  & \multicolumn{1}{c}{$q$}               & \multicolumn{1}{c}{$u$} & \multicolumn{1}{c}{$p$}   & \multicolumn{1}{c}{PA}\\
& &   \multicolumn{1}{c}{UT Date}     & \multicolumn{1}{c}{Start}     & \multicolumn{1}{c}{End}       & \multicolumn{1}{c}{(s)}   & \multicolumn{1}{c}{(nm)}  & \multicolumn{1}{c}{(\%)}  & \multicolumn{1}{c}{(ppm)} & \multicolumn{1}{c}{(ppm)} & \multicolumn{1}{c}{(ppm)} & \multicolumn{1}{c}{($^\circ$)}    \\
\hline
\hline
\parbox[t]{2mm}{\multirow{24}{*}{\rotatebox[origin=c]{90}{\textit{Individual Observations}}}} &
GN  & 2018-07-05    & 12:52:03  & 13:15:54  & 1000  & 463.8         & 93.4  &  $-$48.1 $\pm$ \phantom{0}80.8  &     274.7 $\pm$ \phantom{0}84.5  & 278.9 $\pm$ \phantom{0}82.7   &  50.0 $\pm$ \phantom{0}8.8 \\
&GN  & 2018-07-05    & 13:16:05  & 13:39:33  & 1000  & 463.7         & 93.4  & $-$221.7 $\pm$ \phantom{0}79.6  &  $-$217.5 $\pm$ \phantom{0}77.5  & 310.6 $\pm$ \phantom{0}78.6   & 112.2 $\pm$ \phantom{0}7.4 \\
&GN  & 2018-07-05    & 13:39:59	& 14:02:26  & 1000  & 463.6         & 93.4  & $-$377.4 $\pm$ \phantom{0}83.8  &   $-$18.3 $\pm$ \phantom{0}84.5  & 377.8 $\pm$ \phantom{0}84.2   &  91.4 $\pm$ \phantom{0}6.5 \\
&GN  & 2018-07-05    & 14:02:48  & 14:25:19  & 1000  & 463.6         & 93.4  & $-$338.8 $\pm$ \phantom{0}81.5	&  $-$65.4 $\pm$ \phantom{0}84.3	& 345.1 $\pm$ \phantom{0}82.9	&  95.5 $\pm$ \phantom{0}7.0 \\
&GN  & 2018-07-06    & 13:31:21	& 13:56:08  & 1000  & 463.6         & 93.4  & $-$367.5 $\pm$ \phantom{0}80.0	&  $-$76.0 $\pm$ \phantom{0}83.5	& 375.3 $\pm$ \phantom{0}81.8	&  95.8 $\pm$ \phantom{0}6.4 \\
&GN  & 2018-07-06    & 13:56:37	& 14:20:57  & 1000  & 463.6         & 93.4  & $-$250.6 $\pm$ \phantom{0}79.9	& $-$104.5 $\pm$ \phantom{0}82.9	& 271.5 $\pm$ \phantom{0}81.4	& 101.3 $\pm$ \phantom{0}8.9 \\
&AAT & 2018-08-16    & 14:56:07	& 15:26:47  & 1280  & 451.7         & 63.5  &  $-$27.7 $\pm$ 264.1	&    179.0 $\pm$ 250.7	& 181.1 $\pm$ 257.4	&  49.4 $\pm$ 37.6 \\
&AAT & 2018-08-16    & 15:28:12  & 15:57:43  & 1280  & 451.5         & 63.3  & $-$601.4 $\pm$ 263.5	& $-$181.9 $\pm$ 258.7	& 628.3 $\pm$ 261.1	&  98.4 $\pm$ 13.7 \\
&AAT	& 2018-08-16	& 15:59:01	& 16:28:31	& 1280	& 451.5	        & 63.3	& $-$335.2 $\pm$ 262.4	& $-$153.9 $\pm$ 261.1	& 368.8 $\pm$ 261.7	& 102.3 $\pm$ 25.0 \\
&AAT	& 2018-08-16	& 16:29:56	& 16:59:55	& 1280	& 451.7	        & 63.5	& $-$258.6 $\pm$ 261.7	&    251.6 $\pm$ 243.9	& 360.8 $\pm$ 252.8	&  67.9 $\pm$ 24.7 \\
&AAT	& 2018-08-16	& 17:01:13	& 17:30:56	& 1280	& 451.7	        & 63.5	& $-$224.5 $\pm$ 255.0	&     21.1 $\pm$ 269.3	& 225.4 $\pm$ 262.1	&  87.3 $\pm$ 34.5 \\
&AAT	& 2018-08-16	& 17:32:16	& 18:02:10	& 1280	& 452.0	        & 63.6	& $-$410.9 $\pm$ 264.1	&  $-$51.8 $\pm$ 256.1	& 414.1 $\pm$ 260.1	&  93.6 $\pm$ 22.3 \\
&AAT	& 2018-08-17	& 15:20:43	& 15:50:37	& 1280	& 451.5	        & 63.3	& $-$105.0 $\pm$ 229.5	& $-$304.6 $\pm$ 236.7	& 322.2 $\pm$ 233.1	& 125.5 $\pm$ 25.4 \\
&AAT	& 2018-08-17	& 15:51:58	& 16:22:03	& 1280	& 451.5	        & 63.3	& $-$520.9 $\pm$ 229.5	&     59.3 $\pm$ 243.7	& 524.2 $\pm$ 236.6	&  86.8 $\pm$ 15.3 \\
&AAT	& 2018-08-17	& 16:23:22	& 16:53:40	& 1280	& 451.7	        & 63.5	& $-$333.2 $\pm$ 250.5	&    196.0 $\pm$ 238.1	& 386.6 $\pm$ 244.3	&  74.8 $\pm$ 22.5 \\
&AAT	& 2018-08-17	& 16:55:00	& 17:26:17	& 1280	& 451.7	        & 63.5	&    167.8 $\pm$ 254.9	& $-$190.9 $\pm$ 242.8	& 254.2 $\pm$ 248.9	& 155.7 $\pm$ 31.5 \\
&AAT	& 2018-08-17	& 17:27:42	& 17:57:44	& 1280	& 452.0	        & 63.6	& $-$118.9 $\pm$ 237.8	&    159.3 $\pm$ 243.4	& 198.8 $\pm$ 240.6	&  63.4 $\pm$ 35.2 \\
&AAT	& 2018-08-18	& 16:22:25	& 16:52:12	& 1280	& 451.7	        & 63.5	& $-$488.9 $\pm$ 255.8	& $-$235.1 $\pm$ 231.1	& 542.5 $\pm$ 243.4	& 102.8 $\pm$ 15.1 \\
&AAT	& 2018-08-18	& 16:53:42	& 17:23:19	& 1280	& 451.7	        & 63.5	& $-$283.6 $\pm$ 257.2	&     35.3 $\pm$ 252.0	& 285.8 $\pm$ 254.6	&  86.5 $\pm$ 29.7 \\
&AAT	& 2018-08-18	& 17:24:41	& 17:54:26	& 1280	& 452.0	        & 63.6	&    118.9 $\pm$ 266.0	&    459.8 $\pm$ 251.2	& 474.9 $\pm$ 258.6	&  37.7 $\pm$ 19.2 \\
&AAT	& 2018-08-20	& 16:39:35	& 17:09:58	& 1280	& 451.7	        & 63.5	& $-$198.3 $\pm$ 261.5	& $-$383.2 $\pm$ 241.6	& 431.4 $\pm$ 251.5	& 121.3 $\pm$ 20.7 \\
&AAT	& 2018-08-20	& 17:11:34	& 17:47:50	& 1600	& 452.0	        & 63.6	& $-$764.4 $\pm$ 233.9	&    297.6 $\pm$ 229.4	& 820.3 $\pm$ 231.7	&  79.4 $\pm$ \phantom{0}8.3 \\
&AAT	& 2018-08-21	& 17:32:52	& 18:02:35	& 1280	& 452.2	        & 63.8	& $-$535.4 $\pm$ 259.9	&    211.4 $\pm$ 243.0	& 575.6 $\pm$ 251.5	&  79.2 $\pm$ 14.6 \\
&AAT	& 2018-08-21	& 18:03:57	& 18:33:43	& 1280	& 452.6	        & 64.0	& $-$401.7 $\pm$ 259.7	& $-$140.1 $\pm$ 260.4	& 425.4 $\pm$ 260.1	&  99.6 $\pm$ 21.7 \\
\hline
\parbox[t]{2mm}{\multirow{7}{*}{\rotatebox[origin=c]{90}{\textit{Nightly Means}}}} &
GN	& 2018-07-05	& 12:52:03	& 14:25:19	& 4000	& 463.7	        & 93.4	& $-$237.4 $\pm$ \phantom{0}44.0	&  $-$15.4 $\pm$ \phantom{0}44.6	& 237.9 $\pm$ \phantom{0}44.3	&  91.9 $\pm$ \phantom{0}5.4 \\
&GN	& 2018-07-06	& 13:31:21	& 14:20:57	& 2000	& 463.6	        & 93.4	& $-$308.9 $\pm$ \phantom{0}58.2	&  $-$90.6 $\pm$ \phantom{0}60.5	& 321.9 $\pm$ \phantom{0}59.3	&  98.2 $\pm$ \phantom{0}5.3 \\
&AAT	& 2018-08-16	& 14:56:07	& 18:02:10	& 7680	& 451.7	        & 63.5	& $-$308.8 $\pm$ 106.9	&   17.5 $\pm$ 104.5	& 309.3 $\pm$ 105.7	&  88.4 $\pm$ 10.5 \\
&AAT	& 2018-08-17	& 15:20:43	& 17:57:44	& 6400	& 451.7	        & 63.5	& $-$192.1 $\pm$ 107.2	&  $-$17.7 $\pm$ 107.7	& 193.0 $\pm$ 107.5	&  92.6 $\pm$ 19.7 \\
&AAT	& 2018-08-18	& 16:22:25	& 17:54:26	& 3840	& 451.7	        & 63.5	& $-$226.4 $\pm$ 149.9	&   68.4 $\pm$ 150.0	& 236.5 $\pm$ 145.4	&  81.6 $\pm$ 21.9 \\
&AAT	& 2018-08-20	& 16:39:35	& 17:47:50	& 2880	& 452.0	        & 63.6	& $-$512.8 $\pm$ 174.3	&  $-$25.2 $\pm$ 166.4	& 513.4 $\pm$ 170.3	&  91.4 $\pm$ 10.1 \\
&AAT	& 2018-08-21	& 17:32:52	& 18:33:43	& 2560	& 452.4	        & 63.9	& $-$468.5 $\pm$ 183.7	&   47.8 $\pm$ 177.7	& 470.9 $\pm$ 180.7	&  87.1 $\pm$ 12.3 \\
\hline
\parbox[t]{2mm}{\multirow{3}{*}{\rotatebox[origin=c]{90}{\textit{Means}}}} &
GN	& \multicolumn{3}{c}{2018-07-05 to 2018-07-06}	
                                            & 6000	& 463.6	        & 93.4	& $-$262.4 $\pm$ \phantom{0}37.5	&  $-$40.7 $\pm$ \phantom{0}38.2	& 265.6 $\pm$ \phantom{0}37.8	&  94.4 $\pm$ \phantom{0}4.1 \\
&AAT	& \multicolumn{3}{c}{2018-08-16 to 2018-08-21}
                                            &23360	& 451.7	        & 63.5	& $-$300.4 $\pm$ \phantom{0}59.6	&   13.9 $\pm$ \phantom{0}58.2	& 300.7 $\pm$ \phantom{0}58.9	&  88.7 $\pm$ \phantom{0}5.7 \\
&\multicolumn{4}{l}{Overall (GN + AAT)}                &       &               &       & $-$273.2 $\pm$ \phantom{0}31.7	&  $-$24.3 $\pm$ \phantom{0}31.9	& 274.3 $\pm$ \phantom{0}31.8	&  92.5 $\pm$ \phantom{0}3.3 \\
\hline
\hline
\end{tabular}
\label{tab:pol_obs}
\end{table*}
% End times fixed.

\subsection{Testing for pulsation effects}

Subsequent to the data acquisition it was decided to interrogate the data for evidence of effects resulting from pulsation. As a first step, variability was investigated by calculating the moments of the AAT and GN runs using the methodology of \citet{Brooks94}. The results are presented in Table \ref{tab:stats}. In each instance the calculated kurtosis and skewness (defined so 3 is normal skewness) are not inconsistent with the null hypothesis (of no variability) to a significant level (i.e. 95 per cent). The error variance, which is a measure of whether the standard deviation in the data is consistent with the associated errors, also indicates no variability for the AAT data. This is not surpising as the pulsations have periods much shorter than the length of typical AAT observations, so could be expected to average out, and our errors are of similar magnitude to the mean measurement. However, with the smaller errors associated with the GN data and slightly shorter dwell times, significant non-zero error variances are recorded, which warrants further investigation.

\begin{table}
\centering
\caption{Moment calculations, values are in parts-per-million (ppm).}
\tabcolsep 3 pt
\begin{tabular}{cccccccc}
\hline
\hline
Run & n & Stokes & Mean Err. & Std. Dev. & Err. Var. & Kurtosis & Skewness \\
\hline
AAT & 18 & $q$ & 253.6 & 232.5 & \phantom{00}0.0 & 0.003 & 2.882 \\
    &    & $u$ & 247.1 & 222.1 & \phantom{00}0.0 & 0.001 & 1.977 \\
\hline
GN  &  \phantom{0}6 & $q$ &  \phantom{0}80.9 & 113.8 &  \phantom{0}79.9 & 0.807 & 2.574 \\
    &    & $u$ &  \phantom{0}82.7 & 151.1 & 126.3 & 1.235 & 3.292 \\
\hline
\hline
\end{tabular}
\label{tab:stats}
\begin{flushleft}
Error variance (Err. Var.) is $\sqrt(x^2-e^2)$, where $x$ is the standard deviation (Std. Dev.) and $e$ the mean error (Mean Err.) of a set of measurements.\\
\end{flushleft}
\end{table}

G29-38 has prominent pulsation periods around 800 to 900~s. The total exposure for an observation is usually 1000~s at Gemini North or 1280~s at the AAT, however each observation consists of four measurements at different instrument PAs, with one quarter the exposure. A set of four measurements is necessary to properly remove any instrumental polarization. As a result of triggering overheads, the dwell time for each measurement is about 10 to 20~s longer, meaning that each measurement corresponds to roughly a third of a pulsation cycle. The measurements are separated by about 50 to 100~s owing to Sky measurements. The net effect is that observations at Gemini North and the AAT have dwell times of around 19~min and 25~min respectively (see Table \ref{tab:pol_obs}). 

HIPPI-2 is not photometrically calibrated, and rotating the instrument can produce small jumps in intensity as a result of manufacturing/fabrication tolerances and alignment effects, but intensity (Stokes $I$) is recorded along with Stokes $Q$ and $U$. HIPPI-2's software retains a record of every 2~s integration (which for convenience are grouped together in sets of 10 -- called a repeat in \citealp{Bailey15}). While the arrangement of the observations is far from ideal for this purpose, it is possible to investigate pulsation effects using the intensity and polarization recorded intra-measurement.

\begin{figure}
\centering
\includegraphics[clip, trim={0.5cm 0 1cm 0}, width=8.35cm]{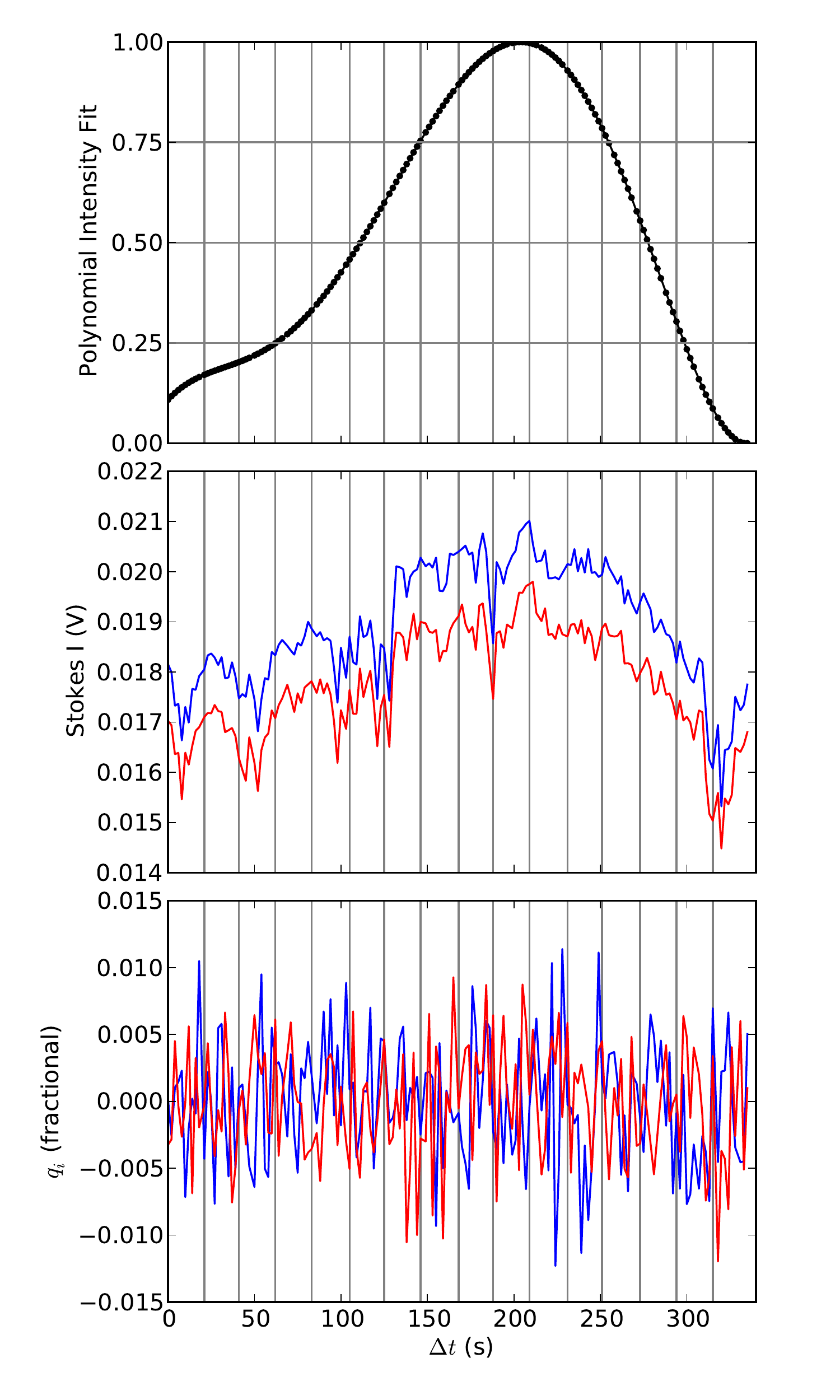}
\caption{An example (AAT) measurement demonstrating the procedure to cut-up the data for pulsation testing. Calculated Stokes parameters are shown individually for Channel 1 (blue) and Channel 2 (red) in the middle (Stokes $I$ in PMT Voltage -- which is linearly proportional to flux in this range, \citealp{pmt}) and lower (Stokes $q$ as fractional polarization). The grey vertical lines divide up the 10-integration sets. In the upper panel is a fifth-order polynomial fit to the average Stokes $I$ data, that has had the minimum value subtracted and then been normalised; the grey horizontal lines correspond to the cut limits.}
 \label{fig:pulse_fit}
\end{figure}

Each measurement has been broken down into its integrations, and a fifth-order polynomial fit to the intensity in that sequence to take out obvious dips from thin clouds or short-duration seeing variability. The fit minimum was then set to zero and the polynomial normalised. This gave an intensity range between 0 and 1 for each integration in each measurement which could be mapped to the polarization measurements. To allow for the calculation of statistical errors in the polarization Stokes parameters the integrations were grouped into their sets of 10. This normalised intensity scale was averaged over an integration set, and then that value used to make a cut of half the data, selecting either the high intensity half (Upper, $I_{Poly}>0.5$), middle (Middle, $0.25>I_{Poly}>0.75$) or the low intensity half (Lower, $I_{Poly}<0.5$). As an example for a single measurement, Figure \ref{fig:pulse_fit} shows the Stokes $I$ parameters calculated in each channel in the middle panel in units of detector voltage; the polynomial to fit the average of the detectors is shown in the upper panel, and the corresponding (instrumental) Stokes $q$ determinations in the lower panel. The vertical grey lines divide up the integration sets. All four PA measurements of the same cut (Upper/Middle/Lower) are then used to calculate polarization values in the standard way. Because each measurement only corresponds to about a third of a pulsation period the upper and lower cuts of the data do not correspond directly to magnitude, but should be indicative of any polarigenic mechanism proportional to magnitude changes.

\begin{figure}
\centering
\includegraphics[clip, trim={0.25 1cm 0.8cm 0}, width=8.4cm]{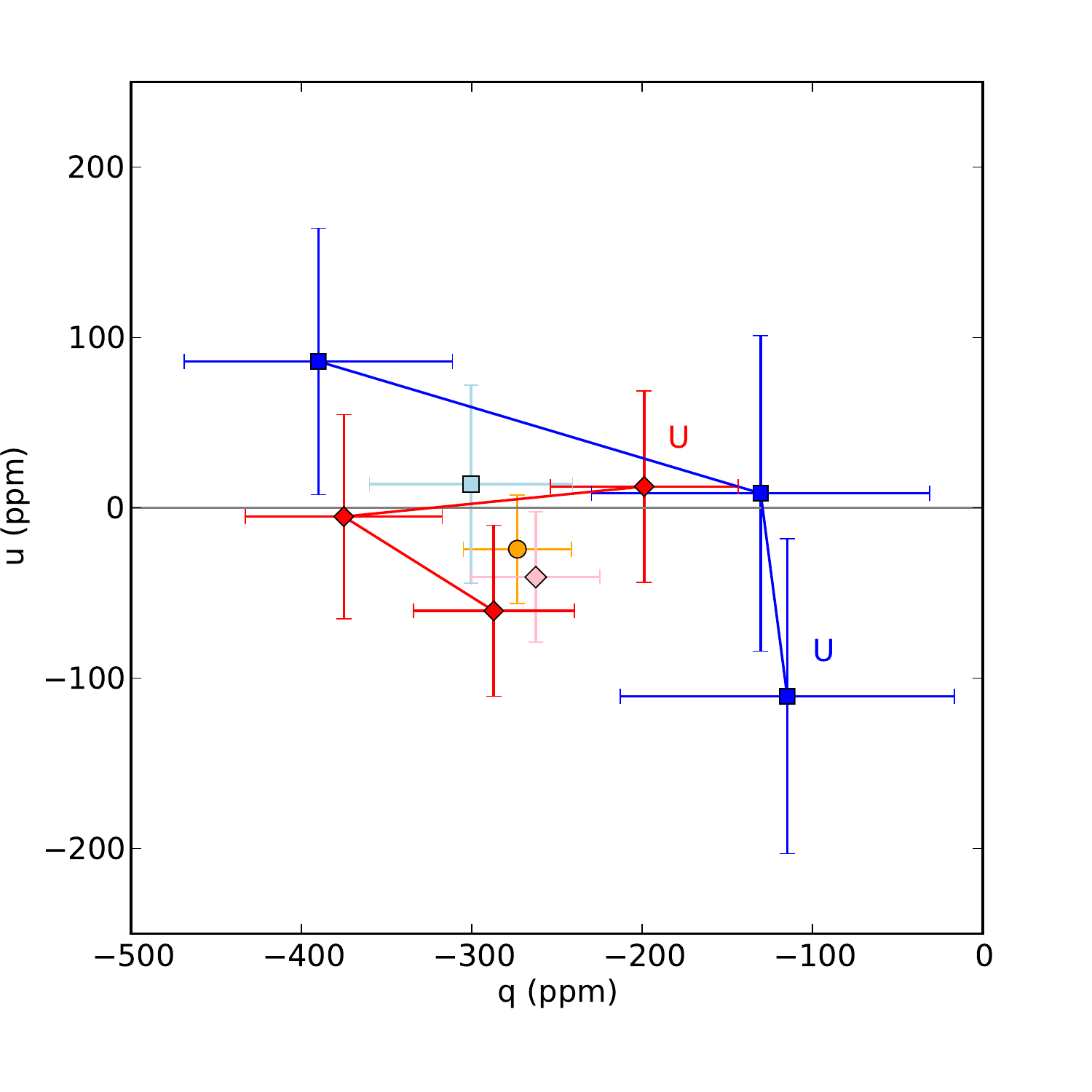}
\caption{Q-U diagrams of cuts of HIPPI-2 observations of G29-38. Gemini North observations are shown as red diamonds, AAT observations as blue squares. The Upper cuts, labelled U, are the first in sequence connected by a thin line to the Middle cuts, and then the lower cuts. The overall mean is shown as an orange circle, and the means of the GN and AAT runs shown as a pink diamond and light blue square respectively.}
\label{fig:pol_cuts}
\end{figure}

There are more significant differences between the various cuts of the data given in Table \ref{tab:pulsation} -- and depicted graphically in Figure \ref{fig:pol_cuts} -- than there are in the nightly means given in Table \ref{tab:pol_obs}. This suggests that there are polarigenic effects resulting from pulsation. Yet, the evidence is inconsistent. The difference between the Upper and Lower cuts of the AAT data are striking in both $q$ and $u$ amounting to 100s of ppm significant to around 2$\sigma$, however the trends are not mirrored in the Gemini data. In the Gemini data it is the Middle and Upper cuts that are most different in $q$, whilst the trend in $u$ is reversed and is not significant.

\begin{table}
\centering
\caption{Cuts by normalised Stokes I.}
\tabcolsep 3 pt
\begin{tabular}{ccrrrr}
\hline
\hline
Run & Cut   & \multicolumn{1}{c}{$q$} & \multicolumn{1}{c}{$u$} & \multicolumn{1}{c}{$p$} & \multicolumn{1}{c}{PA} \\
    &       & \multicolumn{1}{c}{(ppm)} & \multicolumn{1}{c}{(ppm)} & \multicolumn{1}{c}{(ppm)} & \multicolumn{1}{c}{($^\circ$)}    \\
\hline
\hline
GN  & U & $-$198.8 $\pm$ 55.2	&   12.5 $\pm$ 56.2	& 199.2 $\pm$ 55.7	&  88.2 $\pm$ \phantom{0}8.2 \\
GN  & M & $-$375.1 $\pm$ 57.8  &  $-$5.2 $\pm$ 59.9 & 315.1 $\pm$ 58.8  &  90.4 $\pm$ \phantom{0}4.6 \\
GN  & L & $-$287.1 $\pm$ 47.2	&$-$60.5 $\pm$ 50.2	& 293.4 $\pm$ 48.7	&  95.5 $\pm$ \phantom{0}4.9 \\
\hline
AAT & U & $-$114.8 $\pm$ 98.0	& $-$110.6 $\pm$ 92.4	& 159.4 $\pm$ 95.2	& 112.0 $\pm$ 21.2 \\
AAT & M & $-$130.4 $\pm$ 99.3   &    8.5 $\pm$ 92.7  & 130.7 $\pm$ 96.0   &  88.1 $\pm$ 25.8 \\
AAT & L & $-$390.0 $\pm$ 78.7	&   85.9 $\pm$ 78.2	& 399.3 $\pm$ 78.4	&  83.8 $\pm$ \phantom{0}5.7 \\
\hline
\hline
\end{tabular}
\label{tab:pulsation}
\begin{flushleft}
Cuts: U = `Upper', M = `Middle', L = `Lower'.
\end{flushleft}
\end{table}

If one assumes that pulsation effects are present, then the difference between the two runs could either be methodological or phenomenological in origin. The most obvious examples of the latter relate to the pulsations themselves. There are a number of pulsation modes in G29-38 with similar frequencies and beating between those frequencies could result in less prominent effects in one run than another. Although difficult to quantify, the pulsations do appear more prominent in the AAT data. Secondly, G29-38 is an unstable pulsator with modes that come and go, often on timescales of a few months \citep{Kleinman98} and so the star may be behaving differently in one run from the other. Other intrinsic changes in the system could affect the polarization, we examine these in Section \ref{discussion}.

Alternatively, the difference between the two runs could be due to how the data were taken. The shorter duration measurements in the GN run could be inadequately sampling the pulsation cycles for our normalised intensity assignment to be effective -- the smaller the fraction of the pulsation sampled at any given PA, the less appropriate splitting based on intensity \textit{normalised over the exposure} is. An observing program designed to measure distinctly the polarization of the pulsation peaks and troughs is needed to resolve the problem.

\subsection{Interstellar polarization}
\label{interstellar}

\begin{table*}
\caption{Observations of interstellar control stars}
\label{tab:controls}
\tabcolsep 5.25 pt
\begin{tabular}{rlllrrrrrrrrr}
\hline
\hline
\multicolumn{1}{c}{Control}      & SpT   &   Run     & UT                  & Dwell & Exp. & Ap. & $\lambda_{\rm eff}$ & Eff. & \multicolumn{1}{c}{$q$} & \multicolumn{1}{c}{$u$} & \multicolumn{1}{c}{$p$} & \multicolumn{1}{c}{$\theta$}\\
\multicolumn{1}{c}{HD} &        &            &           & \multicolumn{1}{c}{(s)} & \multicolumn{1}{c}{(s)} & \multicolumn{1}{c}{('')} & \multicolumn{1}{c}{(nm)} & \multicolumn{1}{c}{(\%)} & \multicolumn{1}{c}{(ppm)} & \multicolumn{1}{c}{(ppm)} & \multicolumn{1}{c}{(ppm)} & \multicolumn{1}{c}{($^\circ$)}\\
\hline
216385 & F6V & AAT   & 2017-08-18 12:49:58 & 1336 & 960 & 11.7 & 470.3 & 73.9 &   0.8 $\pm$   \phantom{0}8.0 &   $-$16.0 $\pm$   \phantom{0}8.5 &   16.0 $\pm$   \phantom{0}8.2 &   136.4 $\pm$ 17.9 \\
222368 & F7V & AAT-2 & 2019-12-09 10:34:51 &  980 & 640 & 12.9 & 469.2 & 92.9 &   10.5 $\pm$   \phantom{0}5.6 &   $-$26.1 $\pm$   \phantom{0}5.5 &   28.1 $\pm$   \phantom{0}5.5 &  146.0 $\pm$   \phantom{0}5.7 \\
\hline
\hline
\end{tabular}
\begin{flushleft}
Notes: All control star observations were made with the SDSS $g^{\prime}$ filter and the B PMT as the detector. Uncertainties include the positioning error. Spectral types are from \citet{Gray03} for HD~216385 and \citet{Gray01} for HD~222368; the position and distance information presented later is from SIMBAD.\\
\end{flushleft}
\end{table*}

\begin{figure*}
\centering
\includegraphics[clip, trim={2.5cm 0cm 3.4cm 0.95cm},width=17.7cm]{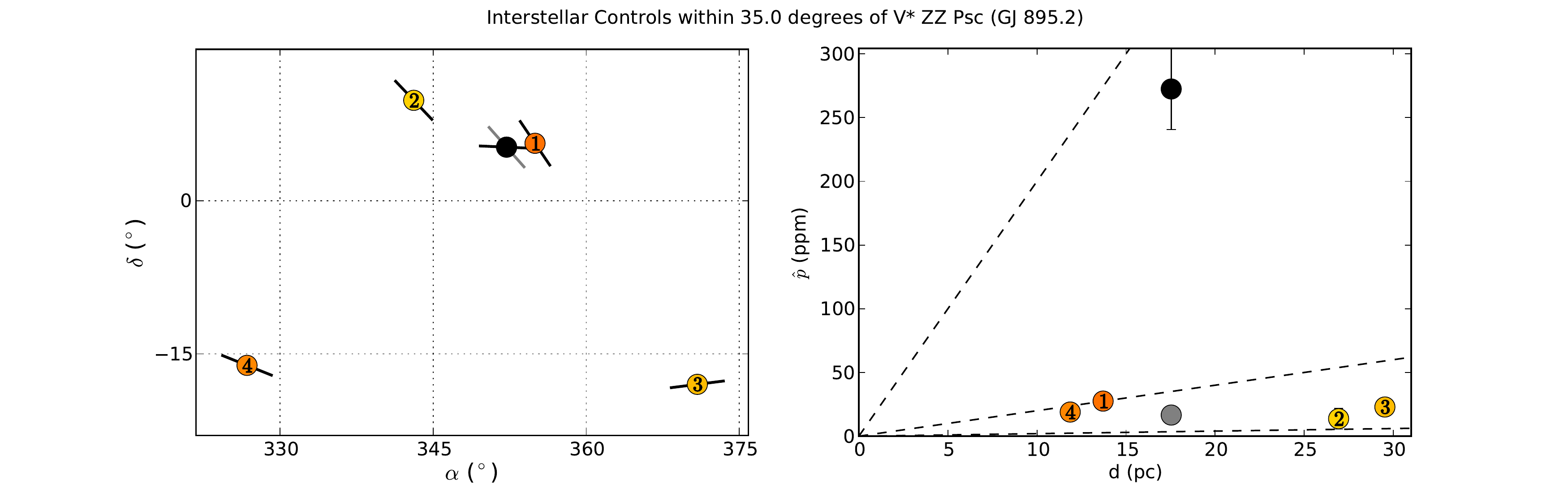}
\caption{A map (left) and magnitude of polarization vs distance ($\hat{p}$ vs $d$) plot (right) of interstellar control stars within 35$^\circ$ of G29-38. Interstellar PA is indicated on the map by the black pseudo-vectors; and defined as the angle North through East, i.e. increasing in a clockwise direction with vertical being 0$^\circ$. The controls are colour coded in terms of $\hat{p}/d$ (where $\hat{p}$ is the debiased $p$) and numbered in order of their angular separation from G29-38; they are: 1:~HD~222368, 2:~HD~216385, 3:~HD~4128, 4:~HD~207098. In the ($\hat{p}$ vs $d$) plot dashed lines corresponding to $\hat{p}/d$ values of 0.2, 2.0 and 20.0 ppm/pc are given as guides. The grey data point is derived from the interstellar model in \citep{Cotton17b} and the black data point represents our best fit interstellar values for G29-38 (converted to 450~nm to best compare with the G29-38 $g^{\prime}$ observations using the Serkowski Law assuming $\lambda_{\rm max}$=~470~nm) after \citet{Marshall16}, which we favour over \citet{Cotton19b} for very nearby objects.}
\label{fig:controls}
\end{figure*}

The measured polarization of G29-38 is a combination of intrinsic and interstellar polarization. G29-38 is within the Local Hot Bubble, very near to the Sun, so its interstellar polarization is expected to be low. The Local Hot Bubble is a region of space carved out by ancient supernovae with a very low gas and dust density that extends from 75 to 150~pc from the Sun \citep{Liu16}. The interstellar medium within this region is not very polarizing, with typical values being around 0.2 to 2~ppm/pc \citep{Bailey10, Cotton16a}, which is an order of magnitude less than the region beyond the Local Hot Bubble \citep{Behr59}. Nonetheless, in consideration of the small polarization magnitude measured for G29-38 in Section \ref{Obs}, the interstellar component may be significant, and here we make an attempt to measure and subtract it.

A common way of gauging the magnitude and orientation of interstellar polarization is to observe nearby intrinsically unpolarized control stars \citep{Clarke10}. We have previously found stars with spectral types ranging from A to early K to be the least intrinsically polarized \citep{Cotton16a, Cotton16b}. Such stars are a good probe of the nearby interstellar medium so long as particularly active stars \citep{Cotton17b, Cotton19a} and those that host prominent debris discs \citep{Cotton17b} are avoided. The dust distribution within the Local Hot Bubble is patchy, making it difficult to measure the true level of interstellar polarization with a control that is not a wide companion. However the ISM is not as patchy within $\sim$25~pc, and we have previously devised a model for the nearest region of space \citep{Cotton17b}. The model relies on a weighted average of nearby control stars with angular separations of less than 35$^\circ$ to determine the position angle, and simple piece-wise linear relationships that vary with Galactic latitude for the magnitude of interstellar polarization. For stars with Galactic co-ordinates $b<+30\degr$, the magnitude of interstellar polarization is given as \begin{equation}\label{eq:p_i} p_i = (1.644 \pm 0.298)(d - 14.5) + (11.6 \pm 1.7),\end{equation} which for G29-38 at $d=$~17.5~pc gives $p_i=16.6~\pm~2.6$~ppm. 

A number of interstellar control stars are to be found in the \textit{Interstellar List} in the appendix of \citet{Cotton17b}. There are two stars in the Interstellar List that are just within 35$^\circ$ of G29-38. So to be able to apply the method from \citet{Cotton17b} we have supplemented the Interstellar List with two new observations reported here in Table \ref{tab:controls} for the first time. The interstellar control stars were observed with HIPPI-2 in the standard way, using the SDSS $g^{\prime}$ filter. The first of these control stars, HD~216385 was observed during the same AAT run as G29-38, and the same TP and PA calibrations apply. The second star was observed some time later in 2019 December at the AAT (denoted AAT-2). During this later run the TP was 20.1~$\pm$~0.7~ppm at a PA of 104.8~$\pm$~1.0$^\circ$ determined from observations of HD~2151, 2$\times$~HD~10700 and 6$\times$~HD~48915; the PA was calibrated from observations of HD~84810 and HD~80558 with a standard deviation of 0.2$^\circ$.

In nearby space it is often difficult to find flawless calibrator stars, this is the case here. Both HD~216385 and HD~222368 have spectral types associated with low intrinsic polarization, but HD~222368 is known to have both hot and cold infrared excesses. The (hot) excess in the K-band is 1.6~per~cent \citep{Nunez17}, however we know that such hot dust is likely made up of nanoscale grains and not very polarizing \citep{Marshall16, Kirchschlager18}. The cold component has a fractional excess of just 1.1$\times10^{-6}$ \citep{Sibthorpe18}, which would only result in a polarization greater than 1~ppm should the disc have an asymmetric geometry. Thus we conclude that HD~222368 is a good interstellar calibrator.

In Figure \ref{fig:controls} it can be seen that the nearest two stars to G29-38 have a similar position angle to each other; this bodes well for the reliability of the interstellar subtraction mechanism. According to \citet{Cotton17b} the position angle is then found by weighting the four stars within 35$^\circ$ as \begin{equation} Wt = (1 - s_a/35),\end{equation} where $s_a$ is the separation to G29-38 in degrees. This gives ${\rm PA}_i=138.5^\circ$; which in turn gives $q_i = 2.0~\pm~1.6$~ppm and $u_i = -16.5~\pm~1.6$~ppm. 

Subtracting this interstellar polarization from that measured for G29-38 gives us a determination for its intrinsic component as $q_\star = -275.2 \pm 31.8$~ppm and $u_\star = -7.8 \pm 32.0$~ppm or alternatively $p_\star = 275.3 \pm 31.9$~ppm (hereafter referred to as $P_{\rm obs}$) at ${\rm PA}_\star = 90.8~\pm~3.8^\circ$.

\section{Analysis}
\label{analysis}

\subsection{Simple models}
\label{basics}

Ignoring for the moment geometrical effects, and other details (such as dust properties etc.), the simple overview of interpreting the observed polarization in terms of scattering from circumstellar dust is as follows:

\begin{enumerate}

\item
From the ratio $f$ of infrared flux $L_{\rm IR}$ to stellar flux $L_{\star}$ \citep{Farihi14},
\begin{equation}
f = L_{\rm IR}/L_{\star} = 0.039,
\end{equation}
we know that the fraction of light from the white dwarf that is intercepted by the dust is $f^{\prime} = f/(1-A_B)$, where $A_B$ is the \textit{Bond albedo} -- defined as the fraction of energy incident on the dust that is not absorbed by the dust\footnote{It is important to note that the word ``albedo'' is used in the literature to mean two different things -- the Bond albedo, as defined here in the text, and the ``geometric albedo'', or ``backscattering albedo'' which is the fraction of incoming radiation that is scattered back in the direction it came from. These two terms are often confused (see the discussion in the review article by \citealp{Hughes18}).}. We shall find that it is likely that $A_B$ is small and hence that $f$ does not differ greatly from $f^{\prime}$.

\item
We conclude that a fraction $f'$ of all photons of a given frequency, which we shall call ``optical'' (e.g.  SDSS $g^{\prime}$ band) strike the dust. Of these photons, a fraction $1-A_B$ are absorbed, and a fraction $A_B$ are scattered. Therefore the fraction of scattered star light is $A_Bf'$.

\item
If the typical polarization of the scattered light is $P_{\rm scat}$ then the observed polarization in the ``optical'' is
\begin{equation}
P_{\rm obs} = P_{\rm scat} \times A_B \times f/(1-A_B).
\end{equation}

\item
For G29-38 we have $f \approx 0.04$. We have found $P_{\rm obs} \approx 0.0003$. From the models below we find a typical value of around  $P_{\rm scat} \approx 0.3$.  From this we see that we expect the approximate value of the Bond albedo to be 
\begin{equation}
A_B \approx 0.025.
\end{equation}

\end{enumerate}

Thus, to summarise, the most direct outcome of our analysis will be an estimate of the Bond albedo $A_B$. However, in order to obtain an accurate estimate we would require both knowledge of the scattering properties of the dust, and of its geometric distribution, both with respect to the star and with respect to the observer; none of this do we have. 

\subsection{Dust properties for modelling}
\label{dust}

Since we do not know what the geometry of the dust is, we shall have to make some plausible assumptions and make do with simple models. It is evident that we shall not be able to elucidate the details of the dust geometry. The main thing that we shall be able to do is to shed light on the dust Bond albedo, and to then compare that with dust elsewhere. Recall that we are interested in the scattering properties at optical wavelengths.

To make a model we shall need some handle on the properties of the dust. In particular we are interested in $A_B$, the scattering phase function $I(\theta)$ which denotes what fraction of the scattered light is scattered through an angle $\theta$, and the polarization function $P_s(\theta)$ which denotes the polarization fraction of the light scattered through an angle $\theta$.

\subsubsection{Scattering phase function $I(\theta)$} 

A commonly used, scattering phase function is that of \citet{Henyey41}. In this case the scattered light intensity varies as 
\begin{equation}
\label{HG}
I(\theta) = \frac{1}{4 \pi} \frac{1 - g^2}{(1 + g^2 - 2 g \cos \theta)^{3/2}},
\end{equation}
which has a single parameter $g$. Here the scattering angle $\theta$ is such that $0^\circ < \theta < 90^\circ$ corresponds to forward scattering. Thus $0 < g < 1$ results in forward scattering and $g = 0$ implies uniform scattering. Observed dust tends to show forward scattering. 

\citet[][figure 6]{Hughes18} show a comparison between this function and observations of several debris discs (as well as Zodiacal dust). Judging from the figure, for these objects, a combination of $g \approx 0.5 - 0.7$, which fits $\theta < 90^\circ$, and $g =0.0 -  0.2$, which fits $\theta > 90^\circ$ might work best.

However, \citet{Krist10} find that for HD~207129, because of the lack of a significant difference between the near and far sides of the ring, they have $0 < g < 0.1$. Similarly, \citet{Golimowski11} find that for the dust in HD~92945, $g = 0.015~\pm~0.015$. \citet{Krist10} contrast this with the tendency of many other discs to show more forward scattering, with $g > 0.15$; these are Fomalhaut $g = 0.2$ \citep{Kalas05}; HD~141569A $g \approx 0.25 - 0.35$ \citep{Clampin03}; AU Mic $g = 0.4$ \citep{Krist05}; HD~107146 $g = 0.3 \pm 0.1$ \citep{Ardila04}. More recently for AU Mic, \citet{Graham07} find a fit with $g = 0.68 \pm 0.01$. In addition, \citet{Ahmic09} in modelling the dust discs in $\beta$ Pic find values of $g = 0.64$ and $g = 0.85$ favoured for their two discs.

We could also note that \citet{Koehler06} report models of light scattering simulations of large cosmic dust aggregates and find values of $g$ in the range 0.2 -- 0.8.

We conclude that overall it seems that using the H-G function (equation~\ref{HG}) is a rough approximation, but the best available. Furthermore, given our paucity of input data for the models it is appropriate to adopt a scattering function with only one parameter. 

\subsubsection{Polarization with scattering angle $P_s(\theta)$}

Here, we consider the question of whether it is sufficient to assume a Rayleigh-like angular variation
\begin{equation}
\label{Rayleigh}
P_s(\theta) = P_{\rm max} \frac{\sin^2 \theta}{1 + \cos^2 \theta}.
\end{equation}

For the debris disc in AU Mic, \citet{Graham07} find that this works with $P_{\rm max} \ge 0.5$. A good fit to the \citet{Graham07} data was obtained by \citet{Shen09}, fitting their models of porous dust grains to the observations, obtaining $g \approx 0.68$ and $P_{\rm max} \approx 0.53$. 

There is more information on the polarization properties of cometary dust and Zodiacal dust. The fits do not seem to be far off the Rayleigh law, except that for scattering angles $150^\circ \le \theta \le 180^\circ$ (i.e. backscattering) the polarization can change sign\footnote{Beware that in many papers the angle $\alpha = 180^\circ - \theta$ is used.}. But since the scattering we are interested in seems to be biased in the forward direction this is not of great concern here. For this dust $P_{\rm max}$ is likely to be in the range 0.1 -- 0.3, with larger values correlating with the presence of silicate features in the spectrum. In addition, \citet{Gupta06} report models of the scattering properties of cometary dust with values of $P_{\rm max}$ in the range 0.2 -- 0.6.

Given all this, we shall consider models using Rayleigh scattering only (equation~\ref{Rayleigh}). We shall use a canonical value of $P_{\rm max} = 0.5$, noting that our conclusions scale simply with this parameter, and will consider primarily two cases, $g=0$ -- corresponding to uniform scattering -- and $g=0.6$ -- corresponding to moderate forward scattering -- as well as values of $g$ in between.

\subsection{Models}
\label{models}

A number of the dust scattering models in the literature apply to objects (standard debris discs) in which the geometry can be resolved. For this reason, many of the models are quite sophisticated, and use dust models which take into account a range of dust particle sizes and properties. 

But in our case, although the dust around G29-38 is often called a ``debris disc'', there is no evidence that the dust is even in the form of an annular disc. The dusty material is assumed to have come from some `planetesimal' or `comet' which formed part of an original debris disc and which has been scattered inwards somehow to get closer to the white dwarf. 

Since we have no knowledge of the actual dust distribution, we consider below some simple ideas that are discussed in the literature (with some slight modifications). \citet{Reach05, Reach09} suggest some idealised models of where the dust around G29-38 might be situated, as well as its mineralogy, in order to fit the infrared spectrum (from 1 to 35~$\mu$m). In the absence of any other information, we use their dust distributions as a starting point.

For the dust, the only parameters we need are the H-G parameter $g$, $A_B$ and $P_{\rm max}$. Everything scales linearly with $A_B$ (provided $A_B \ll 1$) and $P_{\rm max}$. Thus the main variable of interest is $g$.

\subsubsection{Optically thin distribution on a partial sphere}
\label{partsphere}

\citet{Reach09} consider a model with the dust spread in a spherical shell. Since they were trying to fit the infrared spectrum, and therefore need a range of dust temperatures, they need the shell to be physically thick, but optically thin ($\tau_\parallel \approx L_{\rm IR}/L_{\rm opt} \approx 0.039$). \textit{This, being spherically symmetric, would of course give zero polarization in scattered light, which can be ruled out}. Thus, in the first instance, we consider a modified model. Keeping the disc idea in mind, we consider the properties of a section of a spherical shell.  We consider a section of a sphere which (for a spherical polar coordinate system, centred on the star) covers the angles $\pi/2 - \theta_0 \le \theta \le \pi/2 + \theta_0; \; 0 \le \phi < 2 \pi$. This corresponds to a band around the equator ($\theta = \pi/2$) of half-width $\theta_0$. In order to fulfil the infrared flux condition, we would need to increase the optical depth accordingly. And, since we are not here concerned with fitting the infrared spectrum, we can simply assume that the dust shell is infinitesimally thin in the radial direction. For this model, in addition to the dust property parameters $(A_B, P_{\rm max}, g)$, the only geometrical freedom is in $\theta_0$ (the half-width of the band of dust on a sphere), and the inclination angle $i$ between the line of sight and the symmetry axis of the dust ($\theta = 0$).

If the luminosity of the star is $L_*$ and the fraction of absorbed light is $f$, presumed to be given by the infrared excess to stellar flux, with Bond albedo $A_B$, then as above the total luminosity in scattered light is $L_{\rm scatt} = L_* f A_B/(1-A_B)$. If the total polarization magnitude of the band of light (isolated from the star) is $P_{\rm scatt}$ then the polarization magnitude of the band plus star is $P_{\rm obs} = P_{\rm scatt} L_{\rm scatt}/L_* = P_{\rm scatt} f A_B/(1-A_B)$. This leads to:
\begin{equation}
P_{\rm obs}/f = \frac{A_B}{(1-A_B)} P_{\rm scatt} = \frac{A_B}{(1-A_B)} P_{\rm max} P(i,\theta_0 | g) 
\end{equation}
where $P(i,\theta_0 | g)=P_{\rm scatt}/P_{\rm max}$. In this equation, the observables, $P_{\rm obs}$ and $f$ are on the left side. The value of $A_B$ is unknown, and to be determined.  $P_{\rm scatt}$ can be calculated  given the choice of polarization and scattering functions, and the properties of the partial dust shell. We derive $P_{\rm scatt}$  or equivalently $P(i,\theta_0 | g)$ by numerically integrating the distribution of intensity and Stokes parameter $Q$ ($U$ is zero by symmetry and our choice of coordinates) for a band of half-width $\theta_0$ over a range of inclination angles $i$ given a choice of H-G scattering parameter $g$ and maximum polarization $P_{\rm max}$.

\begin{figure*}
\centering
\subfigure{\includegraphics[clip, trim={0 0.5cm 0 0}, width=17.7cm]{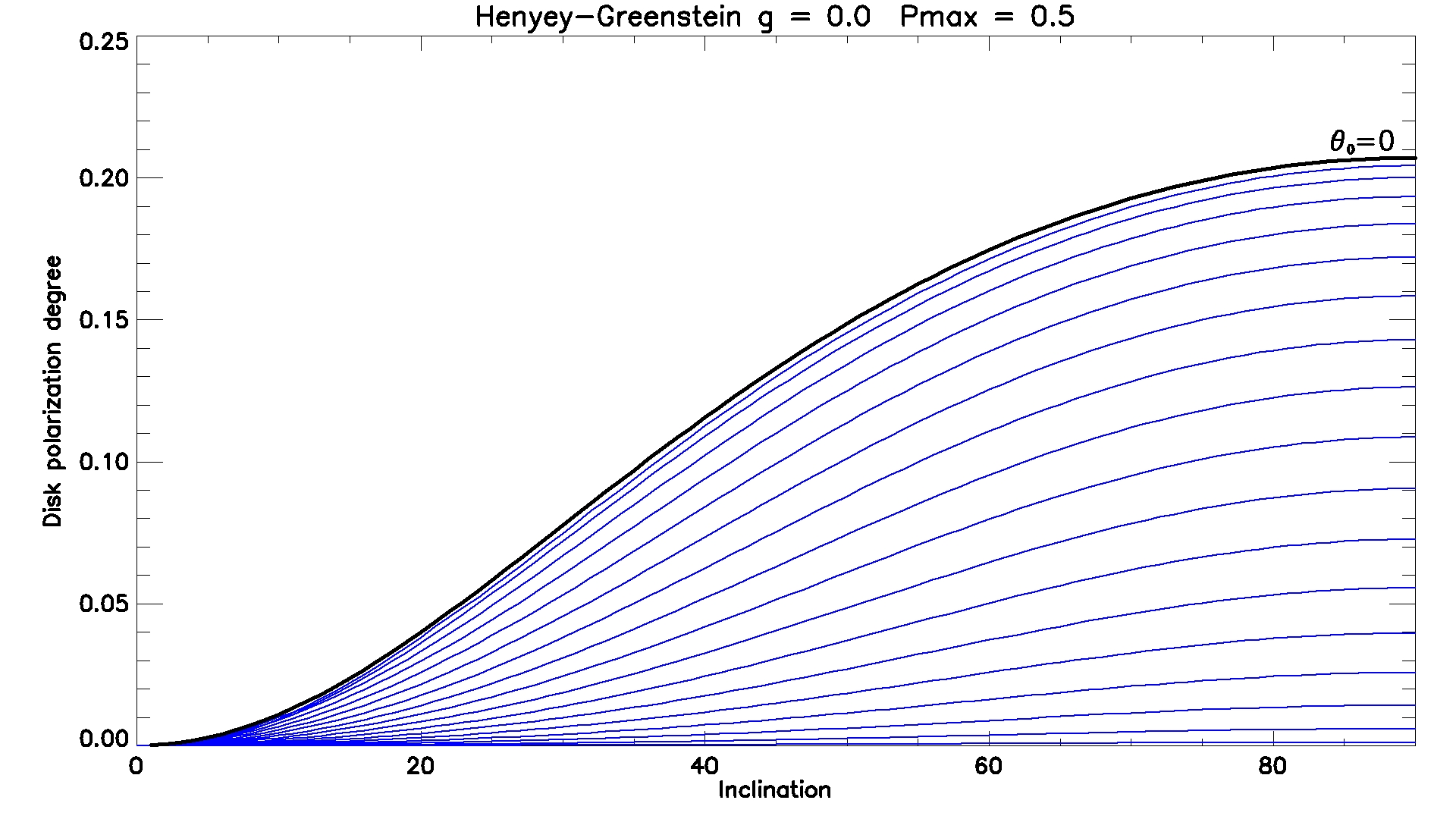}}
\subfigure{\includegraphics[clip, trim={0 0.5cm 0 0}, width=17.7cm]{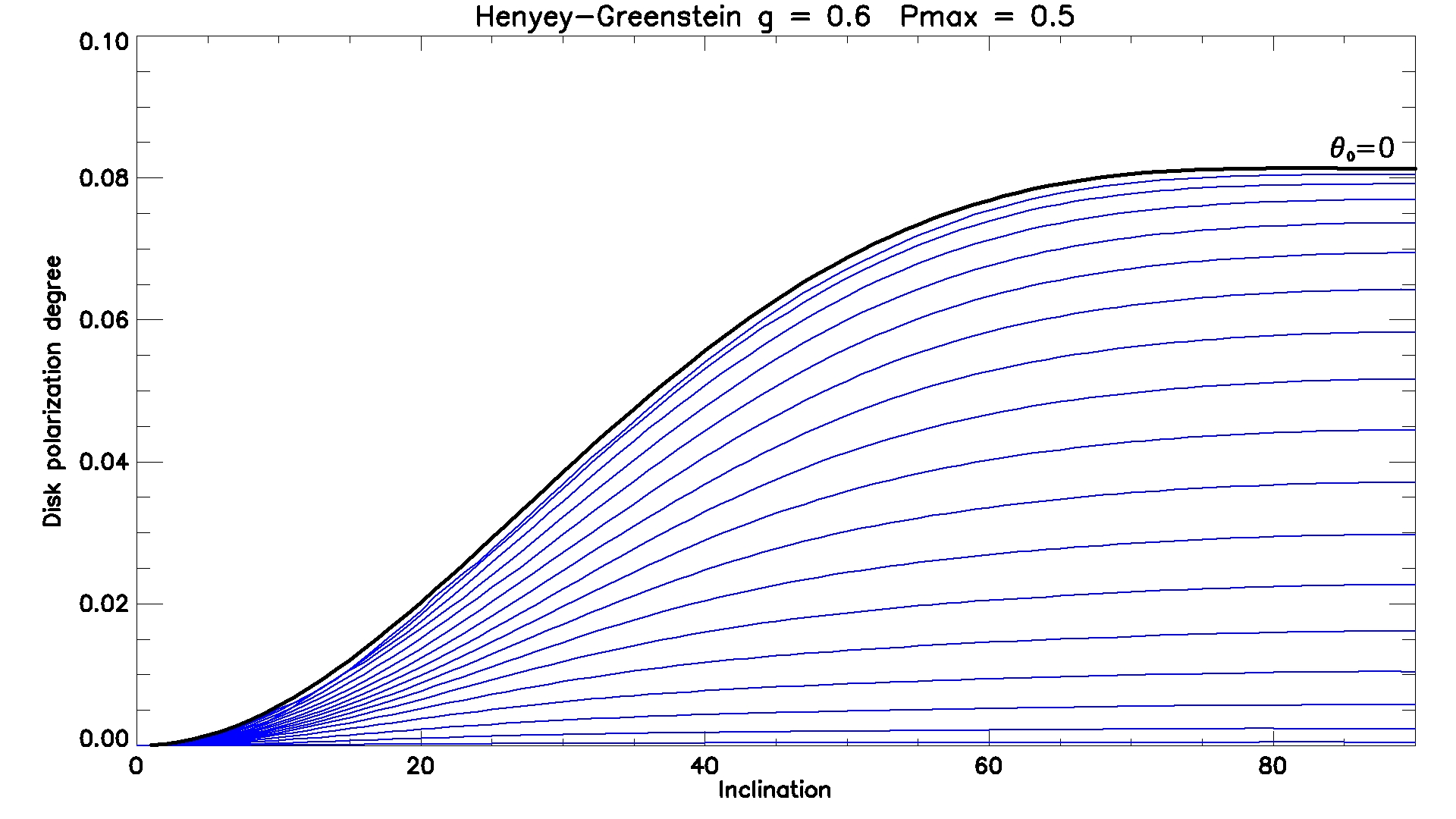}}
\caption{\textit{Top}: Light emitted by a white dwarf is scattered by a surrounding dust shell or disc. The dust is assumed to scatter isotropically $(g = 0)$. The dust is assumed to be in a thin spherical shell, within angles $\pm \theta_0$ around the equator, with angle of inclination $i$ to the observer. Here we plot the degree of polarization of the scattered light. A complete spherical shell would correspond to $\theta_0 = 90^\circ$ and would give zero polarization. The various curves in blue (and annotated) correspond to $\theta_0$ decreasing from $\theta_0 = 86^\circ$ in steps of 5$^\circ$ down to $\theta_0 = 6^\circ$. Large $\theta_0$ corresponds to an almost complete shell, and hence low values of polarization. Inclination $i=90^\circ$, corresponds to an edge-on shell/disc, and thus to maximum polarization. The thick black curve corresponds to the limit of a very thin ring, $\theta_0 \rightarrow 0^\circ$. \textit{Bottom}: The same as in the top panel, but with Henyey-Greenstein parameter $g=0.6$, corresponding to moderately forward scattering dust.}
\label{XX1}
\end{figure*}

Figure~\ref{XX1} shows the isotropically scattered ($g=0$) integrated polarization as a function of inclination for a set of optically thin bands whose half-width is $\theta_0$, with $\theta_0$ running from $86^\circ$ down to $6^\circ$ in steps of $5^\circ$, assuming $P_{\hbox{max}}=0.5$, solved numerically as given in appendix \ref{thinring}. As $\theta_0$ decreases and the bandwidth shrinks, the polarization increases. In the limit, of an infinitely narrow band and isotropic scattering, the polarization magnitude can be solved analytically as shown in appendix \ref{thinring}, plotted here as the thick black curve.  For small inclinations, the shell becomes almost symmetric around the line of sight, and thus the polarization becomes small. Maximum polarization is achieved when the line of sight to the observer lies in the plane of the band (or disc) of dust. For the assumed value of $P_{\rm max} = 0.5$, a typical band-integrated polarization is about 10~per~cent, up to $\sim$20~per~cent for an edge-on thin disc or ring.

For a dust model described by the Henyey-Greenstein scattering phase function with $g=0.6$ -- moderately forward scattering -- the equivalent curves are in the bottom panel of Figure~\ref{XX1}, where the upper envelope is derived numerically from equations (\ref{BillA})---(\ref{BillB}) of appendix \ref{thinring}. The integrated polarization is lower than for isotropic scattering, because of an increased contribution of forward-scattered low-polarization light. Here, for the assumed value of $P_{\rm max} = 0.5$, a typical disc integrated polarization is $\sim$5~per~cent rising to a maximum of $\sim$8~per~cent for the edge-on, geometrically thin case.

\subsubsection{Physically thin disc of dust}
\label{dustring}

The most popular assumption about the geometry of the dust is that it lies in a thin, annular disc (see, for example, \citealp{Rafikov11, Farihi16}). \citet{Reach05, Reach09} consider a thin disc model (which they call the ``physically thick disk'') which is essentially the limiting case $\theta_0 \rightarrow 0^\circ$. Since we know that around 4~per~cent of the radiation from the white dwarf is absorbed and re-radiated by the dust, it is clear that as $\theta_0$ decreases, the optical depth of the dust needs to be increased, becoming around unity when $\sin \theta_0 \approx 0.039$, that is when $\theta_0 \approx 2.23^\circ$. Nevertheless, it is instructive to consider the polarization properties of the limiting case of an optically thin, physically thin disc of dust.

For convenience, we here introduce the parameter $K$ where
\begin{equation}
\label{Kdefinition}
 K \equiv \frac{A_B P_{\rm max} }{1 - A_B}=\frac{P_{\rm obs}}{(f \;  P(i,\theta_0 | g))} .
\end{equation}
For a small value of $A_B$, $K\approx A_B P_{\rm max}$; $P_{\rm obs}$ and $f$ are known from observation and $P(i,\theta_0 | g)$ is calculated theoretically, for a given model.
The observational values are $f\approx 0.039$ (section~\ref{basics}), and $P_{\rm obs}\approx 0.000275$ (sections~\ref{Obs} and \ref{interstellar}).

We see here that the parameter $K$ derives directly from the observations, combined with the properties of the assumed model for the dust. From $K$, given the maximum scattering polarization of the dust ($P_{\rm max}$) we may deduce the Bond albedo, $A_B$.

\begin{figure*}
\centering
\subfigure{\includegraphics[width=17.7cm]{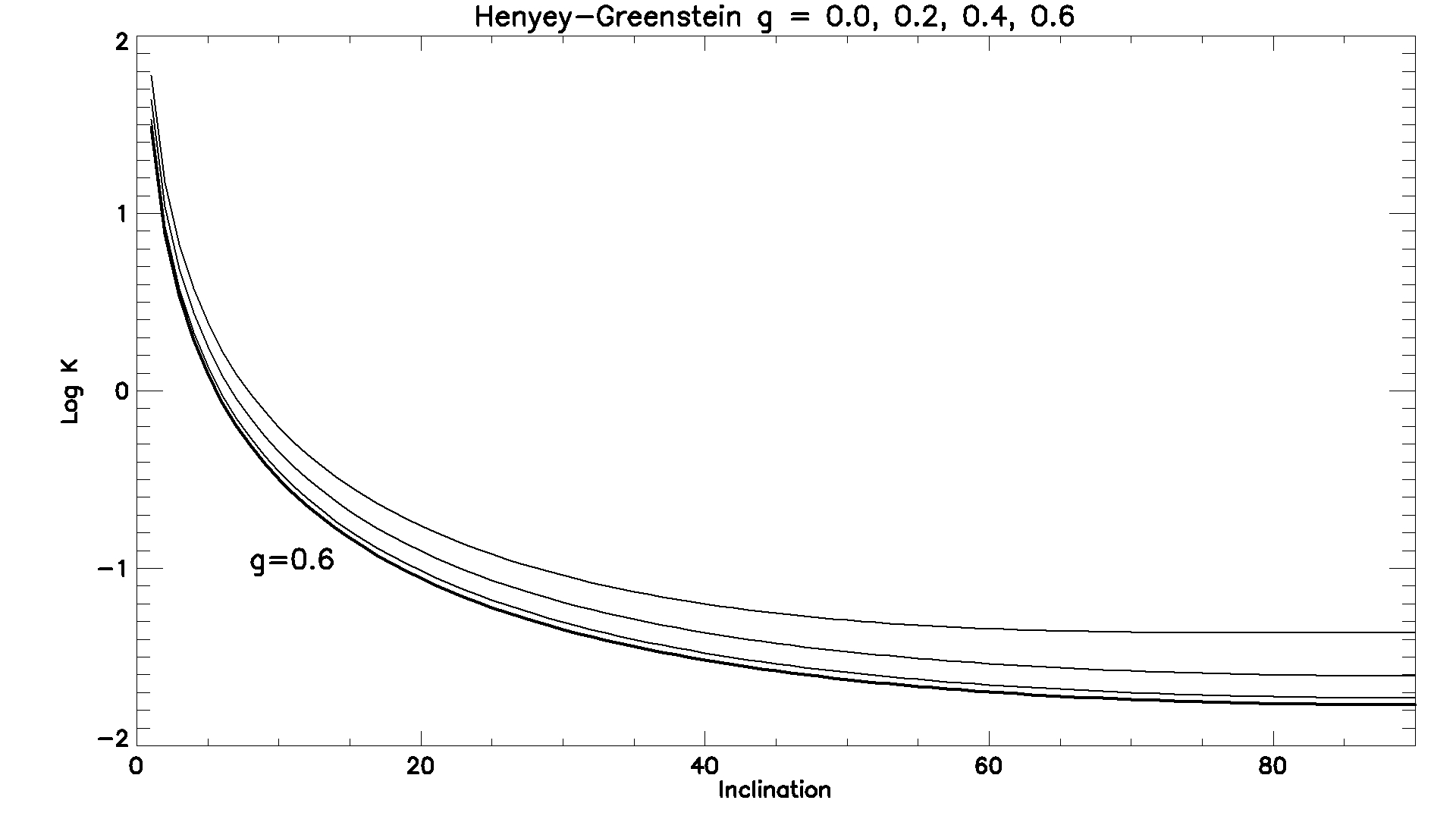}}
\subfigure{\includegraphics[width=17.7cm]{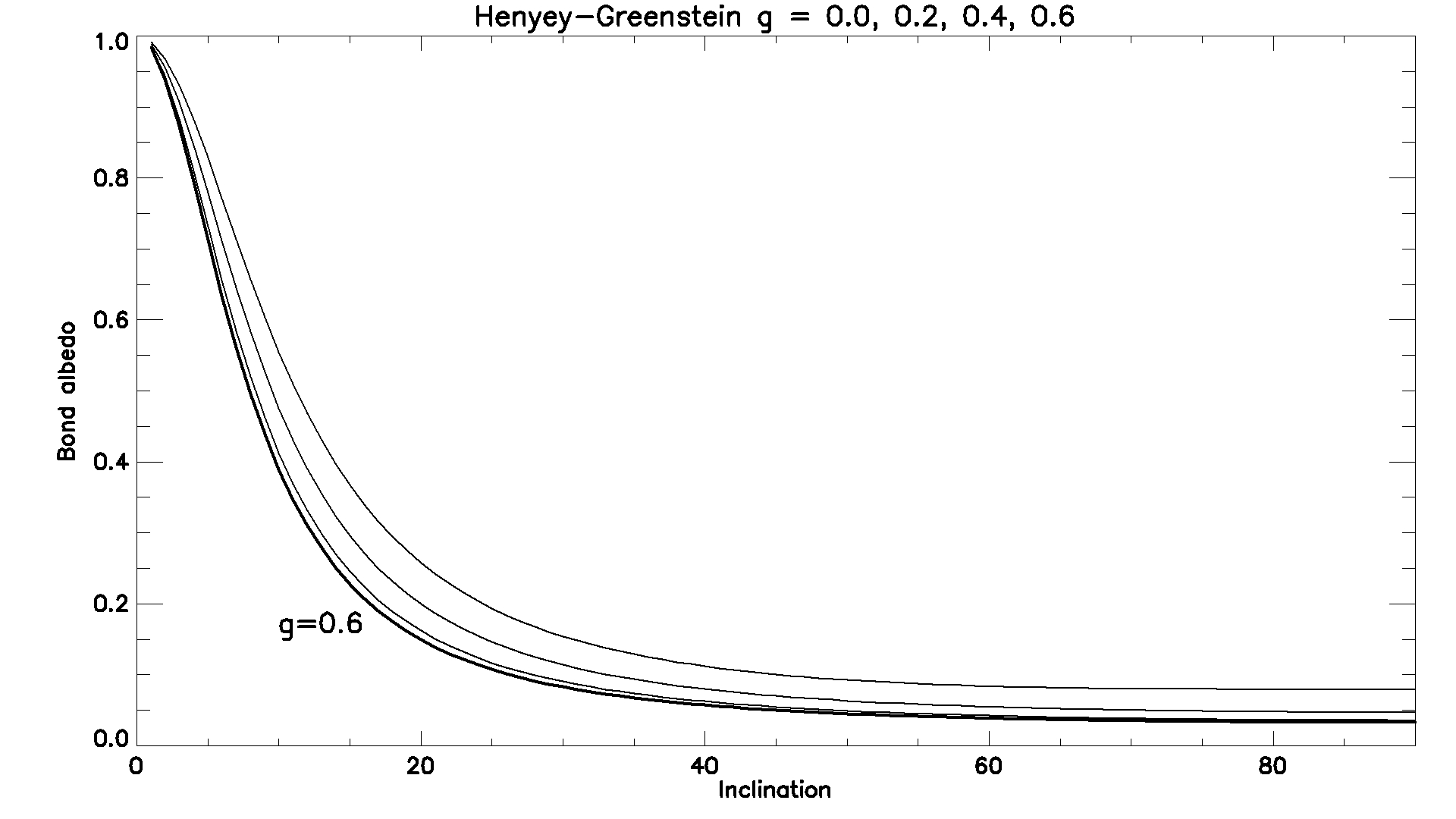}}
\caption{\textit{Top}: The optically thin, physically thin dust disc. The quantity $K$, defined in equation~(\ref{Kdefinition}) is plotted against inclination, $i$, of the disc to the line of sight (with $i=0^\circ$ corresponding to face-on) for different value of the H-G parameter $g = 0.0, 0.2, 0.4$ (thin lines) and $0.6$ (thick line, and annotated) with isotropic scattering ($g=0$) giving the largest values. \textit{Bottom:} For the same discs as in as the top panel, we plot the deduced Bond albedo, $A_B$, as a function of inclination, $i$, for an assumed dust scattering parameter $P_{\rm max} = 0.5$.}
\label{XX3}
\end{figure*}

In Figure~\ref{XX3} we show the functions $K$ and $A_B$ versus inclination for a range of $g$ values derived from the right hand side of equation~(\ref{Kdefinition}), assuming the dust is distributed in a optically thin, geometrically thin disc ($\theta_0 \rightarrow 0^\circ$).

For isotropically orientated discs, the expected value of the inclination is $i = 60^\circ$. At this inclination the value of $K$ ranges from $K\approx 0.020$ to $0.046$ as $g$ ranges from $0.0$ to $0.6$. Assuming $P_{\rm max} = 0.5$, the corresponding values of  $A_B$  are 0.039 to 0.084.

\subsubsection{Optically thick, geometrically thin disc}
\label{soliddisc}

The simple model, discussed for example by \citet{Rafikov11} concerns a model in which an optically thick\footnote{Optically thick in every direction, i.e. radially and in vertical thickness.} layer, but very thin, annular disc of dust orbits the white dwarf. The model is such that $H \ll R_{\star}  \ll R_{\rm disc}$, where $R_{\star}$ is the white dwarf radius, $H$ the disc semi-thickness, and $R_{\rm disc}$ the radius of the inner disc edge. In order for the dust at the inner disc edge to be warm enough to provide the spectrum, it is necessary for $R_{\rm disc}/R_{\star} \approx 10 - 20$ \citep{Reach09, Rafikov11}. 

As far as modelling the polarization is concerned there are two additional complications to the simple optically thin models we considered above. First, because the radiation from the white dwarf strikes the disc surface very obliquely, to a first approximation around one half of the incident flux is scattered into the disc. To a first approximation it might be acceptable to assume that all the light scattered into the disc is absorbed, and thus to replace $A_B$ by $A_B/2$ in equation~(\ref{Kdefinition}) above. However, in reality, a more detailed model requiring some knowledge of the structure of the surface disc layers and then involving polarized radiative transfer is necessary. Second, given the geometry, with $H \ll R_{\star}$, it can be shown (for example by \citealp{Friedjung85}) that the energy received by one side of the disc is (for $R_{\rm disc} \gg R_{\star}$)
\begin{equation}
L_{\rm rec} = \frac{L_\star}{3 \pi} \frac{R_{\star}}{R_{\rm disc}} = 0.106 \; L_\star \frac{R_{\star}}{R_{\rm disc}}.
\end{equation}
The apparent luminosity of the disc (presumably in the infrared, thus $L_{\rm app} = L_{\rm IR})$ then depends on the disc inclination and is given by 
\begin{equation}
L_{\rm app} = \frac{L_\star}{3 \pi^2} \; \cos i \; \left( \frac{ R_{\star}}{R_{\rm disc}} \right) = 0.034 L_\star  \cos i \; \left( \frac{ R_{\star}}{R_{\rm disc}} \right).
\end{equation}
Thus we note that there is a trade-off between increasing the observed polarization (increasing $i$) and the fraction of white dwarf radiation that has to be intercepted by the disc in order to give rise to the observed infrared flux. Given that $ L_{\rm IR}/L_\star = 0.039$, this sets an upper limit to $R_{\rm disc}$ that is too low to be acceptable (even $i=0^{\circ}$ implies $R_\star > R_{\rm disc}$). It is partly for this reason that the model of such a disc by \citet{Reach09} also includes a disc warp.

Given these inherent uncertainties in the disc structure, which would add yet more parameters to any models, it does not seem appropriate at this stage to attempt to draw conclusions for such models from our single polarization measurement.

\subsection{Clumpy dust distribution complications}
\label{clumpy}

Speculatively, if the dust distribution is clumpy, the orbits will complicate an investigation of the pulsations, since the orbital periods may be similar to the pulsation cycles.  While typically an asymmetric dust distribution yields a higher polarization than otherwise expected, the observed polarization will be lower if individual exposures are long enough to integrate over a substantial portion of an orbit.  Uneven dust distributions are observed; for WD~1145+017 \citep{Vanderburg15}, the transits are interpreted as dust orbiting on a timescale of 4.5 to 4.9 hours. This is on the long side for observed dust around white dwarfs. Most estimates of the innermost orbital period of the dust are derived by fitting a series of blackbody profiles to the infrared excess, which for G29-38 and the assumption of a disc morphology \citep{vonHippel07} yields a ring with inner temperature 1150~K and outer temperature 725~K. This in turn yields $R_{\rm dust}$ = 0.15 to 0.28~$R_{\astrosun}$.

Extending to other dust morphologies and dust properties, \cite{Reach09} show that for a spherical dust shell the distance of the innermost dust is weakly dependent on the optical depth, with $R_{\rm dust}$ = 1.3 to 2.6~$R_{\astrosun}$. Taking the mass of the white dwarf as 0.6~$M_{\astrosun}$, the Keplerian orbital period is 752 to 1917~s for the physically thin disc case, which is indeed a similar time scale to the pulsation cycles. For the spherical dust shell, the Keplerian orbital period of the hottest dust is instead 5 to 15 hr.  In the former case, if measured instantaneously, the polarization would be higher than the mean polarization over a whole innermost orbit, and phase curve sampling would have to be considered. For this it would be advantageous to better determine the range of orbital periods, which would ideally require knowledge of both the radial thickness of the dust band and the dust particle size distribution (in particular the minimum dust grain size); the latter is something that might be furnished by multi-band polarimetry.

\section{Discussion}
\label{discussion}

It is useful to compare our results regarding the Bond albedo for the dust around G29-38 to those concerning similar objects. A summary of recent results for the Bond albedo in debris discs is given in the review by \citet{Kimura16}. \citet{Backman92} look at infrared observations and thermal models of the $\beta$ Pic disc. They compare the infrared flux (caused by absorption of stellar flux) with the optical scattering of stellar flux at radii 100~AU $< R < $ 300~AU. They find an albedo for grains in this outer component to be $A_B \approx 0.35$. A later model of the same star is by \citet{Tamura06}, which also uses the previous optical data. Their model uses compact, spherical, optically bright silicate grains whose properties are in accord with the value obtained for the albedo by \citet{Backman92}.

\citet{Krist10} use \textit{Hubble} and \textit{Spitzer} observations of the HD~207129 debris ring. Because they have spatially resolved optical observations of the scattering ring (debris disc, but here a ring with radius $\approx163$ AU and width $\approx 30$ AU) around a G0V star at a distance of 16.0 pc, they can get a good measure of $I(\theta)$ and $P_s(\theta)$. But because they also have infrared data (resolved at 70 $\mu$m), they also have a handle on $A_B$. They find that $A_B \approx 0.05$. From NIR imaging of the young Solar System analog HD~105 \citet{Marshall18} recover values of $A_B$ of 0.15 and 0.06 from \textit{HST}/NICMOS and SPHERE observations respectively. \citet{Choquet18} obtain similarly low scattering albedos from \textit{HST}/NICMOS data for discs around HD~104860 and HD~192748. Such low values for albedo are also seen for asteroids in the Solar System \citep{Morrison77}.

\citet{Krist10} note the work of \citet{Hage90} who model porous grains using 1000 cubes randomly spread through a larger cube. Porosity is defined as the fractional amount of vacuum within the enclosing volume. Thus porosity $= 0$ means solid aggregate, and porosity $= 1$ means vacuum cloud of particles. The highest porosity that they model is around 0.9 -- 0.95, and the lowest albedo ($A_B$) that they get is 0.15 -- 0.3, depending on the material. Thus obtaining for these theoretical models such a low $A_B$ seems to be problematic. \citet{Golimowski11} use \textit{Hubble} and \textit{Spitzer} observations of the debris disc around the nearby K dwarf HD~92945. They find $A_B \approx 0.10$.  They note that standard Mie theory gives values of $A_B \approx 0.55$ from visual to infrared wavelengths. They also note the work of \citet{Hage90}, but contrast it with \citet{Voshchinnikov05} who conclude that albedo \textit{increases} with porosity.

Thus it seems that, within the uncertainties, our results are not out of line with previous findings for the Bond albedo of dust in debris discs, i.e. fairly low values. 

With only a small polarization measured, there is a possibility that some fraction of it might be due to a weak magnetic field. Although the magnetic field of G29-38 has been found to be less than 105~G \citep{Liebert89}, line blanketing has been found to produce 4~ppm/G in the weakly magnetic late-type dwarf $\xi$~Boo A \citep{Cotton19a}. Line blanketing should be a much less effective polarigenic mechanism in a white dwarf where there are fewer spectral lines, however the same mechanism produces around 1.5~ppm/G rotation-modulated amplitude in B-type stars with stronger fields \citep{Wade00}. A similar effect could account for a significant part of the polarization measured if the field is near 100~G. The rotation period of G29-38 is poorly constrained, but there is evidence \citep{Thompson10} that it is approximately 18.5~h.  If so, this is too slow for a significant magnetic field induced polarization to obfuscate pulsation polarization measurements.

Finally we comment briefly on the possibilities afforded by future observations of time-dependent polarization measurements, preferably coupled with synoptic, if not simultaneous, measurements at optical and infrared wavelengths. It is already known that G29-38 displays oscillatory behaviour at optical and infrared wavelengths. In the original analyses by \citet{Graham90, Patterson91}, correlation, as well as lack of correlation, was established between mode frequencies seen at optical and infrared wavelengths. These authors modelled their data in terms of the infrared ($J$, $K_{s}$) variability being caused by re-radiation by a circumstellar ring of dust, of variable flux from the white dwarf. With assumptions about the nature of the modes (in particular that the modes were $l=2$), they were able to put forward ideas about the orientation of the dust disc to the line of sight. However, it is now thought that most of the modes seen in G29-38 correspond to $l=1$ \citep{Kleinman98}. 

Nevertheless it is evident that a more detailed set of observations of the oscillation modes at optical and infrared wavelengths, coupled with polarization measurements, can be used to shed light on models for the circumstellar dust configuration. As a simple example, consider the optically thin dust ring modelled in section~\ref{dustring}. If the central white dwarf is pulsating with an $l=1$ mode, then by symmetry there would be no corresponding oscillatory response seen in the infrared. In contrast, for the optically thick disc/ring, there would be an infrared response to the $m=0$ component of the $l=1$ mode, as measured in the frame of the ring. Also, if the dust scattering were isotropic ($g=0$) there would also, by symmetry, be no variation in the fractional polarization in either case. However, if the scattering is not isotropic ($g \neq 0$) then a variation in the fractional polarization would be observed. 

\section{Summary and Conclusions}
\label{conclusions}

The star G29-38 is a nearby white dwarf which displays a significant infrared excess, corresponding to 3.9 per cent of the white dwarf luminosity. The infrared flux is thought to be caused by absorption and re-emission of white dwarf flux by circumstellar dust. Such dust is also likely to scatter optical photons from the white dwarf photosphere, and such scattered light is likely to be polarized. 

We have presented high precision polarization observations of G29-38 made with two telescopes in the SDSS $g^{\prime}$ band. After correction for interstellar polarization, we measure a polarization $P_{\rm obs}$ = 275.3~$\pm$~31.9~ppm; which is a detection with a significance of 8.6~$\sigma$.

G29-38 also displays significant variability, which is interpreted as being due to non-axisymmetric stellar oscillations \citep{Kleinman98, Winget08}. With this in mind we attempted to look for evidence for such effects in the polarization data. We were unable to detect the effect conclusively, but found enough marginal evidence that such variability might be present. Further investigation of this possibility is warranted, as polarimetric observations with sufficient time resolution might be used to further constrain the dust distribution. In some scenarios, discussed in Section \ref{clumpy}, a clumpy dust distribution would complicate this search.

Our ability to draw conclusions from the current data is limited by our lack of knowledge of the scattering and absorption properties of the dust, and of its geometrical distribution about the white dwarf. Thus (in section ~\ref{basics}) we have made some simple assumptions. We assume that the absorption properties of the dust are described by a single quantity, the Bond albedo, $A_B$ which is the fraction of incoming photons from the white dwarf that are scattered, with the rest ($1-A_B$) being absorbed (and re-emitted as infrared). With regard to the scattering properties, we assume the Henyey-Greenstein form of the scattering phase function (equation~\ref{HG}), with the parameter $g$ on observational grounds taken in the range $g=0$ (isotropic) to $g=0.6$ (moderately forward scattering), and we assume the polarization with scattering angle as Rayleigh scattering (equation~\ref{Rayleigh}), with a maximum value of $P_{\rm max} = 0.5$, again on observational grounds (for details, see section~\ref{dust}). An inherent assumption of all of these models is that the dust is smoothly distributed around the star. If the dust is clumpy as in WD~1145+017 \citep{Vanderburg15, Zhou16} then even geometries that are nominally zero polarization, might produce some small polarizations. Putting this possibility aside, for the geometrical distribution of the dust we draw on ideas based on simple theoretical expectations (e.g. \citealp{Farihi16}), and on ideas used to model the infrared spectrum \citep{Reach09}. 

As a means of illustration we first consider (section~\ref{partsphere}) the dust to be optically thin and to be distributed in an equatorial segment of a spherical shell, of semi-thickness $0 \le \theta_0 \le \pi/2$. In the upper panel of Figure~\ref{XX1} we plot how the polarization expected for such shell segments for isotropic scattering $g=0$, vary with the inclination angle, $i$, of the segments to the line of sight. When $i = 0^\circ$, the segments are viewed along their axes of symmetry and there is no polarization. Maximum polarization occurs when the segments are seen edge on, and the amount of polarization increases as the width of the segments ($\theta_0$) decreases. Maximum polarization occurs when the segments becomes a thin ring ($\theta_0 \rightarrow 0$) and the ring is seen edge-on. In this case the maximum value is $P_{\rm scatt} = 0.21$, for $P_{\rm max} = 0.5$. In the lower panel of Figure~\ref{XX1} we show the corresponding plot when $g= 0.6$, that is for moderately forward scattering. The main point to notice here is that with a stronger degree of forward scattering the maximum polarization, $P_{\rm scatt}$, is significantly reduced, from around 0.2 to around 0.08.

In section~\ref{dustring} we specialise these ideas to the case of the dust lying in a thin ring or disc ($\theta_0 \rightarrow 0$). For the case in which the dust is fully optically thin, for the given dust scattering properties and the observed ratio of infrared excess to stellar flux, $f$, and the observed optical polarization, $P_{\rm obs}$, we are able to obtain the Bond albedo of the dust, as a function of the inclination, $i$. Thus, we find that for most values of the inclination the estimated values of the Bond albedo are around $A_B \approx 0.05 - 0.15$ (although higher values cannot be ruled out if the ring/disc is sufficiently face-on). Such values seem reasonably consistent with estimates of the Bond albedo for debris discs around young stars.

Determining the relationship between the pulsation signal in the optical and infrared offers hope for constraining the geometry. Since the infrared signal is dominated by the dust, if the dust ring is in a face-on inclination there should be a one-to-one relationship between the two wavelength ranges, whereas if the inclination is larger there will be a delay between the signal from the dust behind and in-front of the white dwarf.

We discuss briefly (section~\ref{soliddisc}) the more widely accepted model of an optically thick, but geometrically thin, disc. In this case we conclude that there are too many free parameters to be able to draw credible conclusions.

\section*{Acknowledgements}
We thank the former Director of the Australian Astronomical Observatory, Prof. Warrick Couch, and the current Director of Siding Spring Observatory, A/Prof. Chris Lidman for their support of the \mbox{HIPPI-2} project on the AAT. Nicholas Borsato and Behrooz Karamiqucham assisted with observations at the AAT. We thank Dr. Lucyna Kedziora-Chudczer for useful discussions, and the anonymous referee for their very detailed feedback. 

Based on observations under program GN-2018A-DD-108, obtained at the Gemini Observatory, which is operated by the Association of Universities for Research in Astronomy, Inc., under a cooperative agreement with the NSF on behalf of the Gemini partnership: the National Science Foundation (United States), the National Research Council (Canada), CONICYT (Chile), Ministerio de Ciencia, Tecnolog\'{i}a e Innovaci\'{o}n Productiva (Argentina), and Minist\'{e}rio da Ci\^{e}ncia, Tecnologia e Inova\c{c}\~{a}o (Brazil). This research has made use of the SIMBAD database, operated at CDS, Strasbourg, France. 

Funding for the construction of \mbox{HIPPI-2} was provided by UNSW through the Science Faculty Research Grants Program. TvH acknowledges research support from the National Science Foundation under Grant No.\ AST-1715718. JPM acknowledges research support by the Ministry of Science and Technology of Taiwan under grants MOST104-2628-M001-004-MY3 and MOST107-2119-M-001-031-MY3, and Academia Sinica under grant AS-IA-106-M03.

\bibliographystyle{mnras}
\bibliography{G29-38} 

\appendix

\section{Model Calculation Details}
\label{thinring}

\subsection{Thin ring:  $\theta_0 \rightarrow 0$}

The limiting case for this model occurs when the spherical segment described in Section 5.1 is reduced to a thin ring, 
$\theta_0 \rightarrow 0$. In this case one can obtain some analytic results.

Consider working in Cartesian axes, (X,Y,Z), with the $Z-$axis in the direction of the observer. A ring of radius $R$ tilted at an angle $i$ to the observer is given in these axes as
\begin{equation}
{\bf r} =R  (\cos{\phi}, \cos{i} \sin{\phi}, \sin{i} \sin{\phi}),
\end{equation}
where $\phi$ is the azimuthal angle measured in the plane of the ring, and $\phi = 0$ corresponds the the point ${\bf r} = (1,0, 0)$ in the plane of the sky.

We assume that the white dwarf radius $R_{\star}$ is such that $R_{\star} \ll R$ so that we can treat the white dwarf as a point light source at the centre of the ring. In this case the scattering angle $\zeta(\phi)$ is given by
\begin{equation}
\cos \zeta(\phi) = \hat{\bf r} \cdot \hat{\bf z} = \sin i \sin \phi.
\end{equation}

Then, using the H-G phase function, the scattered intensity is proportional to
\begin{equation}
I(\zeta(\phi)) = \frac{1}{4 \pi} \frac{(1-g^2)}{(1 + g^2 - 2 g \cos \zeta)^{3/2}},
\end{equation}
where $-1 \le g \le 1$.
Then, using a Rayleigh-like polarization scattering function, we find that the polarized intensity of the scattered light is
\begin{equation}
I_P(\zeta) = P_s(\zeta) I(\zeta),
\end{equation}
where
\begin{equation}
P_s(\zeta) = P_{\rm max} \frac{\sin^2 \zeta}{1 + \cos^2 \zeta}.
\end{equation}

To obtain the polarized fraction of scattered light in this model we require the following three quantities:
\begin{equation}
\label{BillA}
Q = \int_0^{2 \pi} I_P(\zeta) \cos 2 \chi \, d \phi,
\end{equation}
\begin{equation}
U = \int_0^{2 \pi} I_P(\zeta) \sin 2 \chi \, d \phi,
\end{equation}
and 
\begin{equation}
I _{\rm tot}= \int_0^{2 \pi} I ({\zeta}) \, d \phi.
\end{equation}
where $\chi$ is the angle on the plane of the sky.
Then the polarized fraction is given by
\begin{equation}
\label{BillB}
P = \frac{\sqrt{ Q^2 + U^2 }}{I_{\rm tot}}.
\end{equation}

We note that for the axes we have chosen, by symmetry, we have that $U = 0$.

{\it Special Case: Isotropic Scattering}

In the case of isotropic scattering, that is $g = 0$ and $I(\zeta) = 1/4 \pi$, we can find analytic expressions for the integrals. We find that $I_{\rm tot} = 1/2$, and that
\begin{equation}
Q(i, g = 0) = \frac{P_{\rm max}}{4 \pi} \int_0^{2 \pi}  \left( \frac{1 - \sin^2 i \sin^2 \phi}{1 + \sin^2 i \sin^2 \phi} \right) \left( \frac{1 - \cos^2 i \tan^2 \phi}{1 + \cos^2 i \tan^2 \phi} \right) \, d \phi.
\end{equation}

This can be written as
\begin{equation}
Q(i, g=0) = \frac{P_{\rm max}}{4 \pi} F(\sin i),
\end{equation}
giving the polarization fraction as
\begin{equation}
P = \frac{P_{\rm max}}{2 \pi} F(\sin i),
\end{equation}
where the function $F(x)$ is given by (Wolfram Research, Inc., Mathematica)
\begin{equation}
F(x) = 2 \pi \left[ 1 + \frac{2}{x^2} \left( \frac{ 1}{\sqrt{1 + x^2}} -1 \right) \right].
\end{equation}
 
 We then find that the polarization fraction varies between $P=0$ for $i = 0$ (face-on) and $P = P_{\rm max} (\sqrt{2}-1) = 0.41 P_{\rm max}$ for $i = 90^\circ$ (edge-on).
 
 {\it General Case: $g \neq 0$}

In this case we have
\begin{equation}
I_{\rm tot} =  \frac{1}{4 \pi}  (1 - g^2) \int_0^{2 \pi} (1 + g^2 - 2 g \sin i \sin \phi)^{-3/2} \, d \phi,
\end{equation}
and
\begin{equation}
\begin{split}
Q(i, g) = \frac{P_{\rm max}}{4 \pi} (1-g^2) \int_0^{2 \pi} & \left( \frac{1 - \sin^2 i \sin^2 \phi}{1 + \sin^2 i \sin^2 \phi} \right) \left( \frac{1 - \cos^2 i \tan^2 \phi}{1 + \cos^2 i \tan^2 \phi} \right) \\
& \times(1 + g^2 - 2 g \sin i \sin \phi)^{-3/2} \, d \phi.
\end{split}
\end{equation}

These integrals can be evaluated numerically using standard techniques.

\subsection{Partial sphere:  $\theta_0 > 0$}
\label{wideband}

For the case where the optically thin band is not narrow, we solved for the scattering and polarization distributions by evaluating them at each pixel individually on a Cartesian grid centered at the white dwarf. Using a pseudo-image approach with dimension up to $4096\times 4096$, we evaluated the scattering angle for each pixel, and hence intensity and polarization contributions with its appropriate weighting, and integrated the results using straightforward summation. For low inclination cases, where more of the light is scattered tangentially, this resulted in some numerical noise as the tangent is undersampled by the pixelation. Nevertheless, in the limiting cases where we can compare the integrations to either the analytic solution for isotropic scattering, or to the numerically solved equations (\ref{BillA})---(\ref{BillB}) above (this Appendix), the upper envelope of polarization given by these limiting cases is in good agreement with the numerical approach.

%%%%%%%%%%%%%%%%%%%%%%%%%%%%%%%%%%%%%%%%%%%%%%%%%%

% Don't change these lines
\bsp	% typesetting comment
\label{lastpage}
\end{document}